\documentclass[fleqn,10pt]{wlscirep}
\usepackage[utf8]{inputenc}
\usepackage[T1]{fontenc}

\usepackage{amsmath,amssymb}

\usepackage{changepage}

\usepackage{textcomp,marvosym}

\usepackage{cite}

\usepackage{nameref,hyperref}

\usepackage[right]{lineno}

\usepackage{array}

\newcolumntype{+}{!{\vrule width 2pt}}

\usepackage{dirtytalk}
\usepackage{booktabs}
\usepackage{subcaption}
\usepackage{annotate-equations}

\definecolor{color_1}{HTML}{009E73} %
\definecolor{color_2}{HTML}{E03A3D} %
\definecolor{color_3}{HTML}{0072B2} %
\definecolor{color_4}{HTML}{000000} %
\definecolor{color_5}{HTML}{F0E442} %
\definecolor{color_6}{HTML}{CC79A7} %
\definecolor{revisions}{HTML}{000000}

\usepackage{soul}
\colorlet{soul_1}{color_1!20}
\colorlet{soul_2}{color_2!20}
\colorlet{soul_3}{color_3!20}
\colorlet{soul_4}{color_4!20}
\colorlet{soul_5}{color_5!20}
\colorlet{soul_6}{color_6!20}

\colorlet{local_load}{soul_5}
\colorlet{visits}{soul_2}
\colorlet{inters}{soul_3}
\colorlet{eod}{soul_1}

\newcommand{\highlight}[2]{\sethlcolor{#1}\hl{#2}}

\soulregister{\highlight}{2}

\newcommand{\supplementarysection}[1]{%
  \setcounter{figure}{0}%
  \let\oldthefigure\thefigure%
  \renewcommand{\thefigure}{S\oldthefigure}%

  \setcounter{table}{0}%
  \let\oldthetable\thetable%
  \renewcommand{\thetable}{S\oldthetable}%

  \setcounter{subsection}{0}
  \let\oldthesubsection\thesubsection
  \renewcommand\thesubsection{S\arabic{subsection}}

  \newpage
  \section*{#1}%
}

\usepackage{lastpage,fancyhdr,graphicx}
\usepackage{epstopdf}

\title{%
Simulating nationwide coupled disease and fear spread in an agent-based model} %

\author[1,2,*]{Joy Kitson}
\author[2,3]{Prescott C. Alexander}
\author[4]{Joseph Tuccillo}
\author[3]{David J. Butts}
\author[3]{Christa Brelsford}
\author[1]{Abhinav Bhatele}
\author[5]{Sara Y. Del Valle}
\author[2,+]{Timothy C. Germann}
\affil[1]{Department of Computer Science, University of Maryland, College Park, Maryland, 20742, USA}
\affil[2]{Theoretical Division, Los Alamos National Laboratory, Los Alamos, New Mexico, 87544, USA}
\affil[3]{Analytics, Intelligence and Technology Division, Los Alamos National Laboratory, Los Alamos, New Mexico, 87545, USA}
\affil[4]{Geospatial Science and Human Security Division, Oak Ridge National Laboratory, Oak Ridge, Tennessee, 37830, USA}
\affil[5]{Associate Laboratory Directorate for Global Security, Los Alamos National Laboratory, Los Alamos, New Mexico, 87545, USA}

\affil[*]{jkitson@umd.edu}
\affil[+]{tcg@lanl.gov}

\begin{abstract}
    Human cognitive responses, behavioral responses, and disease dynamics co-evolve over the course of any disease outbreak, and can result in complex feedbacks. We present a dynamic agent-based model that explicitly couples the spread of disease with the spread of fear surrounding the disease, implemented  within the EpiCast simulation framework. EpiCast  models transmission across a realistic synthetic population, capturing individual-level interactions. In our model, fear propagates through both in-person contact and broadcast media, prompting individuals to adopt protective behaviors that reduce disease spread. In order to better understand these coupled dynamics, we create and compare a range of \textcolor{revisions}{compartmental models to ensure that introducing additional disease states does not prevent the emergence of multiple waves in these simpler models}. Additionally, we compare a range of behavioral scenarios within EpiCast, varying the level and intensity of fear and behavioral change. Our results show that the addition of asymptomatic, exposed, and pre-symptomatic disease states can impact both the rate at which an outbreak progresses and its overall trajectory \textcolor{revisions}{in compartmental models. In EpiCast}, the combination of non-local fear spread via broadcasters and strong behavioral responses by fearful individuals generally leads to multiple epidemic waves, an outcome that occurs only within a narrow parameter range when fear spreads purely through local contact. Accounting for the coupled spread of fear and disease is critical for understanding disease dynamics and designing timely, targeted responses to emerging infectious threats.
\end{abstract}


\begin{document}

\flushbottom
\maketitle
\thispagestyle{empty}

\section{Introduction}

Looking back on the early stages of the COVID-19 pandemic reveals a dizzying array of responses across all levels of society, across the entire world. The response to the COVID-19 pandemic may have been the largest coordinated behavior change in the history of humanity. From state-mandated lockdowns and school closures to individual masking, testing, and vaccination, these responses shaped both personal experiences of the pandemic, and also observed outcomes at a societal level ~\cite{chernozhukov2021causal,navas2022forecasting,toharudin2021national,suthar2022public}.

Unfortunately, accurately anticipating individual behavior has proved challenging for the similarly diverse array of models developed to simulate the disease's spread~\cite{cramer_evaluation_2022}. The most common method of accounting for behavioral changes has been to use exogenous data sources, such as cell phone mobility data or historical public health policies, to impose top-down behaviors~\cite{nixon2022evaluation}. However, exogenous approaches struggle to capture the two-way feedback between behavioral changes and disease spread, since they mostly treat behavior as an input parameter, able to be scheduled in advance~\cite{bauch2013behavioral}. Truly dynamic behaviors require incorporating a behavioral model into the heart of a simulation, able to respond flexibly to changes in the global or local simulation state.

However, most such \emph{endogenous} behavioral models operate within high-level compartmental models where a series of ordinary differential equations (ODEs) govern the changing proportion of the population in different disease states~\cite{hamilton2024incorporating, schluter_unraveling_2023}. Such models can only answer population-level questions about disease spread because they smooth over the heterogeneity of real-world populations. In contrast, agent-based models (ABMs) capture this heterogeneity by directly modeling the disease state and behavior of each individual in the simulated population. This capability has enabled the use of ABMs for retrospective analysis of COVID-19 spread patterns~\cite{murakami2022agent}, evaluation of public health interventions~\cite{germann2006mitigation,kersting2021predicting,bosman2024agent}, and the study of differential COVID-19 outcomes across regions and demographic groups~\cite{chen2024role}. However, ABMs rarely include endogenous behaviors~\cite{hamilton2024incorporating}, and the few existing ABMs which do only simulate relatively small populations consisting of at most a million people~\cite{mao2014modeling}.

To address these gaps, we present a \emph{coupled contagion} model of interdependent disease and fear spread implemented in EpiCast. EpiCast is a production ABM capable of simulating the spread of an infectious disease -- such as COVID-19 -- on a \textcolor{revisions}{digital twin} of the full US population while accounting for varying demographic and occupational distributions, as well as commute and flight patterns across the country~\cite{germann2006mitigation,alexanderEpicast20Largescale2025}. In our model, fear of a disease spreads alongside the disease itself, influencing individuals to adopt various protective behaviors that either reduce their number of contacts or the likelihood they are infected by any given contact. This fear can spread through either in-person contact -- much like the disease -- or through broadcast media. To our knowledge, this is the first work to model the coupled dynamics of two contagions in an agent-based simulation of this scale and complexity.

Additionally, we present a series of ODE-based models we developed to \textcolor{revisions}{ensure that solely introducing new disease states used in EpiCast would not disrupt the dynamics found in existing ODE-based coupled contagion models}. EpiCast includes several disease states beyond the standard susceptible, (symptomatic) infectious, recovered (SIR) model, such as asymptomatic and presymptomatic infectious states. As most prior coupled contagion models have been based on the SIR model, \textcolor{revisions}{we found it useful to first} adapt existing ODE-based models to account for these new disease states, \textcolor{revisions}{settling on a high-level design} before translating the resulting system into the agent-based context. We then analyzed how the incremental inclusion of disease states in these \textcolor{revisions}{ODE} models changes their dynamics.

{\color{revisions}
    In the process of designing and evaluating our model, we sought to answer the following research questions:
    \begin{enumerate}[start=1,label={\bfseries RQ \arabic*}]
        \item How does the introduction of additional disease compartments change outbreak dynamics?
        \label{rq:compartments}
        \item How do different types of fear-induced behavior impact outbreak dynamics in a large scale ABM?
        \label{rq:scenarios}
        \item How do key parameters in our models relate to important outbreak characteristics, such as the total attack rate or the number of epidemic waves?
        \label{rq:params}
    \end{enumerate}
}

\section{Related work}
\textcolor{revisions}{The most common mechanism for incorporating changing behavior into epidemic models is using data-driven, \emph{exogenous} behavior models. Exogenous models treat behavior as an independent variable, rather than a simulation output; it cannot be tightly coupled with simulation results like the prevalence of the disease at a given time. In contrast, \emph{endogenous}} behavior models incorporate a framework for human behavior directly into a simulation. \textcolor{revisions}{For example, students could stay home from school after their teacher tests positive for a disease. This tight coupling allows endogenous models to} react flexibly to novel scenarios. \textcolor{revisions}{Such} models have proven able to produce multiple epidemic waves and at times dramatically improve model predictions~\cite{rahmandad2022enhancing,tovissode2024heterogeneous}, and generally fall into three categories~\cite{hamilton2024incorporating}:
\begin{enumerate}
    \item \emph{Feedback loop models}, where behavior directly depends on disease prevalence. \textcolor{revisions}{This can be in the form of either agents adopting behavior when the disease prevalence passes a certain threshold, or by scaling adoption probabilities or intensities based on prevalence.} One noteworthy example by Rahmandad et al. outperformed the majority of the U.S. Center for Disease Control COVID-19 Forecasting Hub models by scaling the transmission rate in a SEIR compartmental model by a prevalence-dependent scaling factor~\cite{rahmandad2022enhancing}. Kasa et al. provide an overview of such models~\cite{kassa2011epidemiological}.
    \item \emph{Game theoretic models}, where individuals select an option that maximizes a utility function describing their preferred outcomes. \textcolor{revisions}{These models can typically consider two types of \emph{agents}, individual decision-makers with their own behaviors: 1) central planners, which select policies or incentive structures to maximize a global measure of social good, and 2) self-interested agents seeking to maximize their private utility. these types of agents may be used along or together.} Huang et al. survey models of this type~\cite{huang2022game}.
    \item \emph{Coupled contagion models}, where an individual's behaviors depends on their beliefs. \textcolor{revisions}{An individual's beliefs are generally represented as discrete compartments, similar to disease states, with transition probabilities determined by} a combination of other individuals' beliefs and information regarding the state of the pandemic. \textcolor{revisions}{Beliefs and prevalence information can both be either global or local, spreading between neighbors in an ABM, and therefore may} spread in a manner similar to the disease. Epstein et al. present such a model where two contagions -- fear of the disease and the disease itself -- spread simultaneously through a population~\cite{epstein2008coupled}, adding a third contagion -- fear of vaccination -- in a later version of the model~\cite{epstein2021triple}.
\end{enumerate}
Several models also combine elements from multiple categories above. Poletti et al. developed a behavioral model for H1N1 Influenza where individuals choose between protective and normal behaviors so as to maximize a utility function based on the recent disease prevalence~\cite{poletti2011effect}. Jovanovic et al. compare models of vaccine uptake using an objective function which incorporates population-level prevalence information and local-level information spread through neighbors in a information diffusion network~\cite{jovanovic2021modelling}.

\subsection{Agent-based modeling of endogenous behavior}
Despite the range of developed behavioral models, the vast majority are built on top of standard compartmental models, most of which use a simple feedback loop. In particular, there are only a handful of agent based models (ABMs) which use coupled contagion models~\cite{hamilton2024incorporating}.

Epstein et al.~\cite{epstein2008coupled} develop a relatively simple ABM where 1800 agents, move around a 2D grid. Agents can spread fear and disease to agents in adjacent cells, and can either flee, isolate or retain the same behavior while fearful. Note that while they demonstrate the emergence of waves in the ordinary differential equation (ODE) version of their model, they fail to do so for the ABM. Rajabi et al.~\cite{rajabi2021investigating} also use a grid-based ABM, but only have agents quarantine when fearful. They demonstrate how multiple waves can emerge in their simulation and simulate the impact of travel restrictions and contact tracing on disease spread in a 5000 agent population. \textcolor{revisions}{Both ABMs are highly idealized, with agents lacking realistic schedules.} Palomo-Briones et al.~\cite{palomo2022agent} present a more complex but still spatially explicit 400 agent model where agents have a short visit schedule consisting of visits to their home, social activities, and economic activities, and can chose to wear masks, social distance, or wash hands based on a combination of other agents' beliefs and the results of their previous behavior. Curiel et al.~\cite{prieto2021vaccination} use a small 5000-agent network-based ABM where a disease spreads alongside anti-vaccine beliefs under different network topologies and vaccination strategies. \textcolor{revisions}{While these two models introduce more sophisticated behaviors, they deal with very small populations, much like the previous models.} Mao~\cite{mao2014modeling} presents a detailed simulation of the Buffalo metropolitan area with nearly a million agents, with a schedule of activities drawn from demographic data. Agents can decide to use flu prophylaxis using a threshold-based decision model based on either infection risk or prophylaxis adoption among contacts. While these models range in complexity, only Rajabi et al., with the simplest model, demonstrates an ability to produce multiple epidemic waves. In addition, the largest simulated population is smaller than most U.S. states and tuned for a narrow geographic area, which limits the usefulness of its results for state-level and national policy-makers.

\subsection{The EpiCast simulator}
The smaller scales of the coupled contagion ABMs described above contrast sharply with that of HPC-scale simulators like EpiCast. EpiCast simulates disease spread on a massive artificial surrogate population of the full U.S. containing over 322 million agents~\cite{germann2006mitigation,alexanderEpicast20Largescale2025}. In the baseline version of EpiCast, this behavior relies on a combination of fixed exogenous responses based on demographics and top-down public health interventions taken as input. The population used in EpiCast is built using UrbanPop~\cite{tuccillo2023urbanpop}, a framework which uses American Community Survey (ACS) Data to create high-fidelity synthetic populations. Compared to similar frameworks, UrbanPop captures more of the heterogeneity in real-world populations, producing cross-sectional representations of agents across sociodemographic, economic, educational, and mobility characteristics. UrbanPop assigns each member of the generated population to a specific household in a specific census block group, in a manner which ensures the aggregate demographics of the generated block group correspond precisely to those observed in census data. In EpiCast, these households form the basis for dividing the population into communities, within which agents have varying contact intensities defined by whether they share a household, neighborhood, workplace, school, etc. These contact intensities determine the pairwise likelihood that disease transmission occurs between two agents within the same community. Agents then commute between a home and a work or school community each day in a manner consistent with Department of Transportation commute flow data, and occasionally engage in long distance travel to geographically distant communities based on flight data.

\section{Methods}
In designing our model, we seek to adapt an existing coupled contagion ODE model to function within EpiCast, which requires accounting for all of the disease states present in the original code. We first construct a SIR-based ODE-model as a simple baseline, then incrementally incorporate new disease states until we have accounted for all those used by EpiCast. Once this design is complete, we then modify the model to function in an individualized context. In the process, we introduce both short- and medium-distance mechanisms for fear spread and two types of behavioral responses to fear with an eye toward reproducing the twin epidemic waves found in the ODE models.

\subsection{Coupled contagion ODE models}
We take a streamlined version of Epistein et al.'s model of coupled fear and disease spread~\cite{epstein2008coupled} as our starting point due to its flexibility and ability to produce multiple epidemic waves. We then incrementally add the asymptomatic, presymptomatic and exposed disease states present in EpiCast to this model\textcolor{revisions}{, in order to ensure that the dynamics found in the simpler models are not disrupted by the addition of these disease states. The final model uses all disease states present in EpiCast.} We refer to this series of models using the Cartesian product of the two types of states. For example, we describe a standard $SIR$ model paired with a fear model with \textbf{N}eutral and \textbf{F}earful states as an $SIR \times NF$ coupled model. \textcolor{revisions}{Note that these models are not intended to be directly compared with the extension of EpiCast described in Section~\ref{sec:epicast-fear}. We instead use them to inform the high-level design of the model implemented in EpiCast, with a focus on how to couple the dynamics of disease, fear, and behavior.}

We designed the following series of models:
\begin{enumerate}
    \item $SIR \times NF$, the closest to Epstein et al.'s original model. The flow rate from neutral to fearful compartments is slower by a factor of $\rho_f$ from the \textbf{R}ecovered state. $\rho_f = 0$ corresponds to the Epstein model.
    \item $SI_sI_aR_sR_a \times NF$, distinguishing between \textbf{s}ymptomatic and \textbf{a}symptomatic \textbf{I}nfectious and \textbf{R}ecovered states. Of the disease states, only $R_s$ has a reduced flow rate from neutral to fearful compartments \textcolor{revisions}{and contributes to a reduction in fearful agents. This is because, given that we do not currently model testing, agents are only aware of symptomatic infections}.
    \item $SEPI_sI_aR_sR_a \times NF$, adds an \textbf{E}xposed state to represent the incubation period and a \textbf{P}resymptomatic state to allow for disease spread prior to the potential onset of symptoms. This is the closest to the model implemented in EpiCast.
\end{enumerate}
For brevity, we present only the $SEPI_sI_aR_sR_a \times NF$ system in full. The $SEPI_sI_aR_sR_a \times NF$ flow diagram is shown in Supplementary Fig. S1 and the full system of equations is given in Supplementary Section S1. \textcolor{revisions}{The other models are special cases of this system and can be derived by setting the the compartments not present in a given model to 0.}

\subsection{Coupled contagion agent-based model}
\label{sec:epicast-fear}
The EpiCast coupled contagion model can be thought of as a translation of the $SEPI_sI_aR_sR_a \times NF$ model into the individualized context of EpiCast. This requires reworking aspects of the basic functionality of the original EpiCast simulator to facilitate fear spread and tie an individual's behavior to their fear state.

\subsubsection{Modeling fear spread}
\label{sec:fear}
We implement two methods of fear spread in EpiCast: one which operates based on an agent's local neighborhood of contacts in a manner similar to disease spread, and a second that approximates the effect of traditional newspaper publishers and broadcast media. For simplicity, we do not distinguish between these two forms of media, and refer to media outlets of either type as \say{broadcasters} below when describing the model.

In order to implement local fear spread, we took advantage of EpiCast's preexisting methodology for computing disease spread, refactoring the code to allow for fear spread using different per-contact transmission probabilities. Fear is spread by symptomatic and fearful agents and countered by agents who know they have recovered from an infection and have a neutral fear state (i.e. agents in disease state $R_s$ and fear state $N$).

For non-local fear spread, we created broadcasters capable of taking one of three positions on the disease. They may either: (1) spread fear, (2) counter fear spread, or (3) remain neutral. In order to site these broadcasters, we begin by examining the occupations of agents in each community, which are identified by 2017 North American Industry Classification System (NAICS) codes in EpiCast. If we identify at least one agent who works in publishing (NAICS code 511) or broadcasting (NAICS code 515) in a given community we locate a media outlet there and assign all agents within the relevant industry to work there. Each broadcaster is able to potentially influence anyone within the census tract in which they are located, and starts by taking a neutral position.

Once initialized, a broadcaster's position is determined by a combination of the fear states of its employees and the level of disease spread in the broadcast area. When the proportion of employees at a broadcaster who are fearful of the disease crosses a certain threshold, $p_\text{bc\_start}$, the broadcaster begins spreading fear. We then record the current infection rate, the number of new infections in the broadcast area as a proportion of its population, as $\dot{E}_0$. In future timesteps, the broadcaster determines its position by comparing the current infection rate, $\dot{E}_t$, against $\dot{E}_0$ as follows:

\begin{enumerate}
    \item If $\dot{E}_t < p_\text{bc\_counter} \cdot \dot{E}_0$, the broadcaster begins countering fear spread.
    \item Otherwise, if $\dot{E}_t < p_\text{bc\_neutral} \cdot \dot{E}_0$, the broadcaster takes a neutral position.
    \item Otherwise, the broadcaster spreads fear of the disease.
\end{enumerate}
where $p_\text{bc\_counter} < p_\text{bc\_neutral}$.

For each agent in the broadcast area of a given broadcaster, they have a fixed chance, $p_\text{bc}$, of being influenced by that broadcaster in a given timestep if the broadcast is not neutral and their fear state does not correspond to the broadcaster's position.

\subsubsection{Modeling behavior}
\label{sec:behavior}
We modeled two different ways in which agents may change their behavior when they are fearful: (1) taking protective measures (such as masking and hand-washing) while maintaining their normal routine, and (2) withdrawing from their normal routine and remaining at home. We model this first measure by reducing the susceptibility of agents in state $<S, F>$, similar to how this is handled by Epstein et al.~\cite{epstein2021triple} and in the ODE model.

In the second case, we modify an agent's routine so that they remain in their home community, neither commuting to their work community nor engaging in long distance travel, and only interacting with other members of their household. If currently traveling, they immediately return to their home community. Note that this is different from a complete quarantine, as they can still transmit the disease or their fear to their household. If agents are fearful, they can either withdraw purely due to fear with a fixed probability $p_\text{fear}$ for a given duration $t_\text{with}$, or due to having a symptomatic infection (i.e. being in state $<I_s,F>$). Note that any agent with a symptomatic infection will withdraw if their symptoms are severe enough for them to be hospitalized, regardless of their fear state.

\textcolor{revisions}{\subsection{Experimental setup}}

First, \textcolor{revisions}{we sought to examine how different sets of disease compartments impact outbreak dynamics (\ref{rq:compartments}). We find numerical solutions to each ODE model an initial value problem (IVP) solver -- the \texttt{solve\_ivp} function from the Scipy integrate package (v1.14.0) -- under different parameter choices. In the process,} we identify scenarios under which each of our ODE models produce multiple epidemic waves, providing a comparison of when the same parameter values produce substantially different outcomes between models. Parameter values are shown in Supplementary Table~S1.

Second, we sought to \textcolor{revisions}{compare the outbreak dynamics resulting from several different sets of fear-induced behaviors (\ref{rq:scenarios}). We run a range of scenarios in EpiCast where agents withdraw from everyday activities under certain conditions (most in part due to fear) and has a reduced susceptibility to infection due to protective actions. This includes withdrawls by agents who are 1) hospitalized agents (hospitalized withdrawals), 2) symptomatic and fearful (sick withdrawals), and 3) only fearful (fear-only withdrawals). These scenarios also compare the impact of purely} local (based on physical contact with agents in the same community) and non-local (based on broadcasters operating within a given census tract) fear spread modes. For these scenarios, we use a synthetic population of about 322 million agents representing the 48 contiguous U.S. states and Washington DC. We seeded index cases on a per-county basis using estimates of active cases on 25 March 2020 derived from data from The New York Times, based on reports from state and local health agencies. For the first five days of the simulation, for an average of 856 thousand seed cases per day (about 0.265\% of the population). Parameter values are shown in Supplementary Table~S2.

Next, \textcolor{revisions}{we investigated how a couple key model parameters impact the dynamics produced by the behavioral model implemented in EpiCast (\ref{rq:params}). We conduct} a series of sensitivity analyses comparing EpiCast outputs for a range of values of $p_\text{fear}$ and $\sigma_f$, which determine the strength of an agent's behavioral response to fear of the disease. We draw these values from the range $[0, 1]$, trying five values for each parameter for a total of 25 combinations. We run each combination twice -- once with only local fear spread and once with both local and broadcaster-based fear spread. These experiments are conducted on the contiguous U.S. dataset described above. Choices for parameters other than $p_\text{fear}$ and $\sigma_f$ were held fixed during these experiments and correspond to Scenarios (c) and (d) from Supplementary Table~S2 (or, equivalently, Scenarios (e) and (f)) without and with broadcasters, respectively. In order to compute the number of epidemic waves programmaticly, we first compute a centered 7-day rolling maximum of new case counts -- so as to avoid identifying peaks caused solely by weekly variation -- then call the \texttt{find\_peaks} function from the Scipy signal package (v1.14.0) in Python with a prominence of $0.0001$. This value indicates that new cases should fall by at least 1 new infection per 10,000 people before rising again if we are to consider a given local maximum to be the peak of a given epidemic wave. We find this value avoids spurious peaks (i.e. those in the long tail of some outbreaks) while still capturing those most visually prominent in the data.

Finally, we also conducted two supplementary experiments to explore the impact of stochasticity, initial conditions (namely the pattern of initial infections), and the threshold of fearful workers at which broadcasters begin spreading fear ($p_\text{bc\_start}$). These experiments were conducted on a smaller population of $\sim5.6$ million people representing the U.S. state of Colorado. Similarly to the sensitivity analysis experiments conducted on the full U.S., all other parameter values were held fixed and correspond to Scenarios (c) and (d) from Supplementary Table~S2. The results of these experiments are discussed in Supplementary Section~S3.

\textcolor{revisions}{\section{Results}}
\label{sec:results}
\textcolor{revisions}{We describe the results of the three main sets of experiments described above.}

\subsection{Comparing ODE models}
\label{sec:compartments}

\begin{figure}[t]
    \centering
    \begin{subfigure}{0.33\textwidth}
        \centering
        \includegraphics[width=\textwidth]{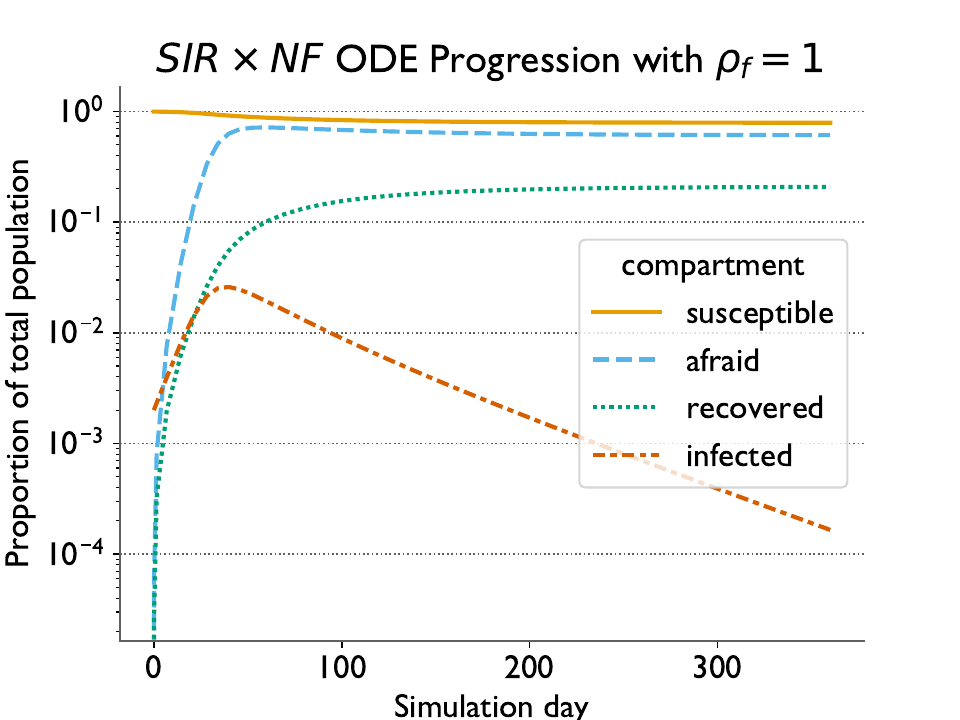}
        \caption{}
        \label{fig:sir_rho_1}
    \end{subfigure}
    \begin{subfigure}{0.33\textwidth}
        \centering
        \includegraphics[width=\textwidth]{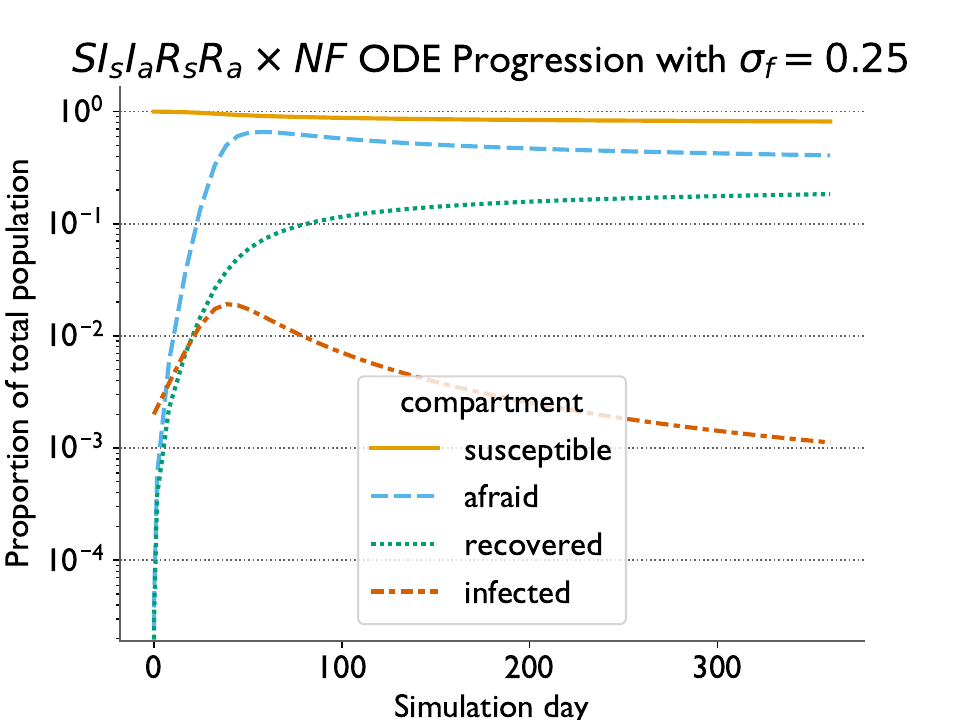}
        \caption{}
        \label{fig:si_ai_sr_sr_a_sigma_f_25}
    \end{subfigure}
    \begin{subfigure}{0.33\textwidth}
        \centering
        \includegraphics[width=\textwidth]{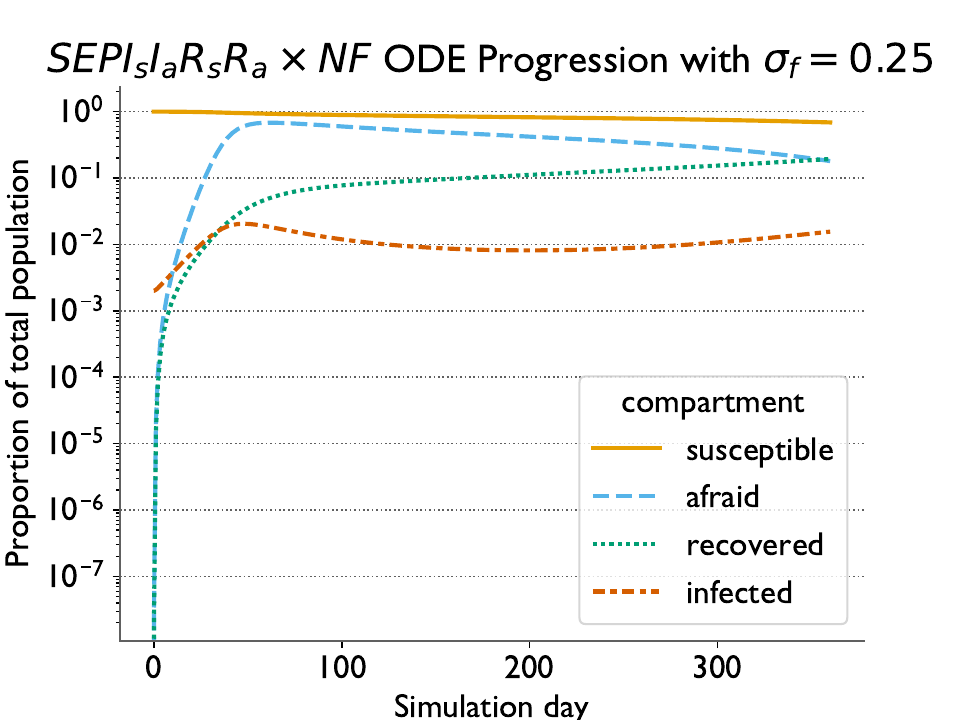}
        \caption{}
        \label{fig:sepi_ai_sr_sr_a_sigma_f_25}
    \end{subfigure}

    \begin{subfigure}{0.33\textwidth}
        \centering
        \includegraphics[width=\textwidth]{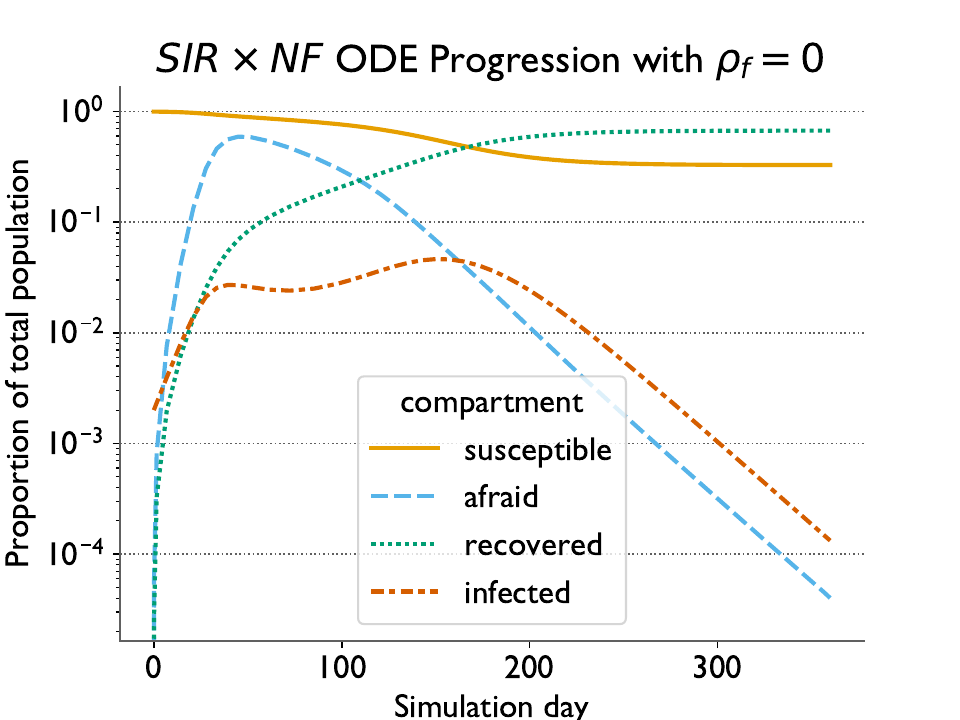}
        \caption{}
        \label{fig:sir_rho_0}
    \end{subfigure}
    \begin{subfigure}{0.33\textwidth}
        \centering
        \includegraphics[width=\textwidth]{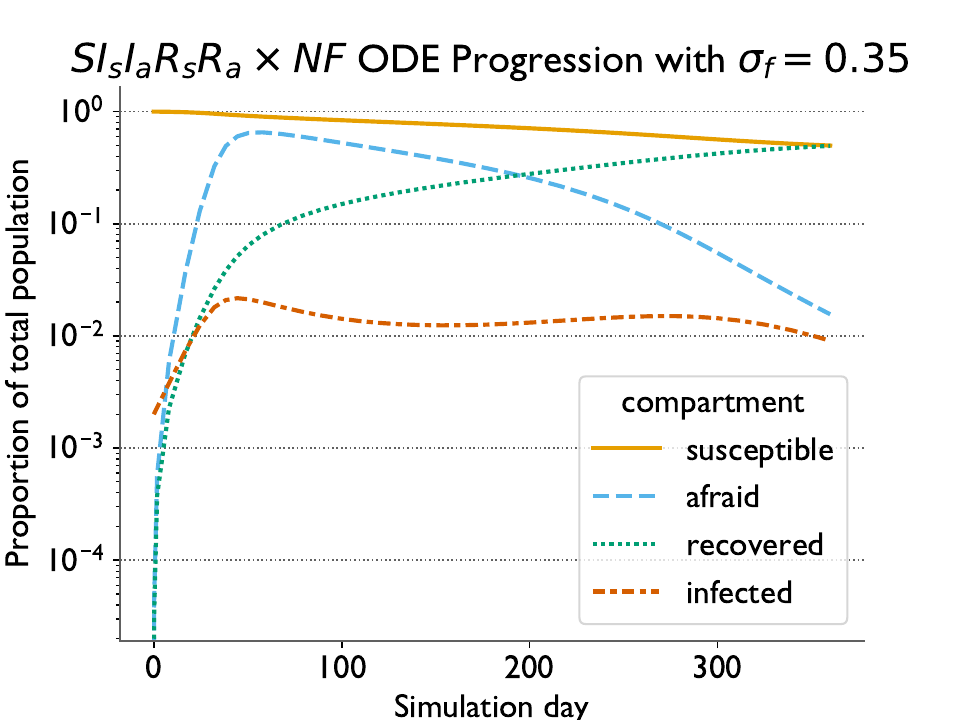}
        \caption{}
        \label{fig:si_ai_sr_sr_a_sigma_f_35}
    \end{subfigure}
    \begin{subfigure}{0.33\textwidth}
        \centering
        \includegraphics[width=\textwidth]{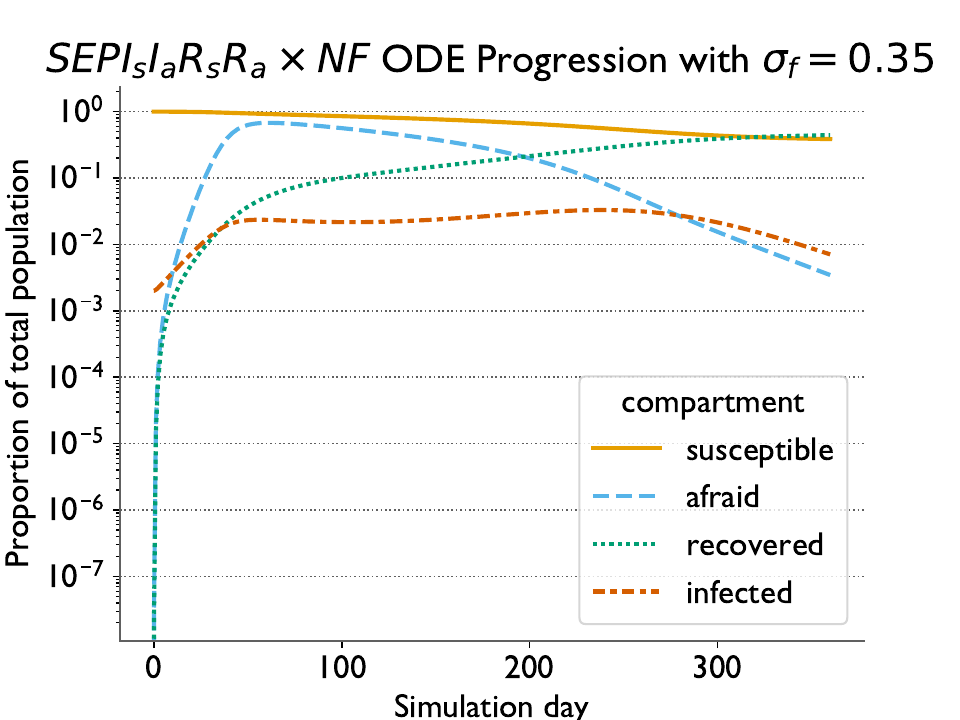}
        \caption{}
        \label{fig:sepi_ai_sr_sr_a_sigma_f_35}
    \end{subfigure}
    \caption{{\bf ODE model comparison:} We show outputs with attention to parameter choices which impact the emergence of multiple epidemic waves. On the top and bottom we show solutions corresponding to parameter choices which fail to produce, or produce, respectively, a second wave within 360 days. We show these results for \textcolor{revisions}{models combining 1) a basic \textbf{S}usceptible \textbf{I}nfectious \textbf{R}ecovered model with \textbf{N}eutral and \textbf{F}earful states ($SIR \times NF$, \ref{fig:sir_rho_1} and \ref{fig:sir_rho_0}, varying the relative likelihood an individual becomes fearful after covering from a symptomatic infection, $\rho_f$), 2) seperating \textbf{s}ymptomatic and \textbf{a}symptomatic infectious and recovered states ($SI_sI_aR_sR_a \times NF$,\ref{fig:si_ai_sr_sr_a_sigma_f_25} and introducing \textbf{E}xposed and \textbf{P}resymptomatic states (\ref{fig:si_ai_sr_sr_a_sigma_f_35}, varying the susceptibility of fearful individuals to infection, $\sigma_f$), and $SEPI_sI_aR_sR_a \times NF$ (\ref{fig:sepi_ai_sr_sr_a_sigma_f_25} and \ref{fig:sepi_ai_sr_sr_a_sigma_f_35}, varying $\sigma_f$)}. For ease of comparison, we combine all infected (including $E$), recovered, and fearful compartments in each model into a single line in each of the plots above.}
    \label{fig:ode-results}
\end{figure}

For the ODE comparison experiments, Fig.~\ref{fig:ode-results} shows the solutions corresponding to the parameter choices that either failed to produce a second epidemic wave within 360 days (top) or produced such a wave (bottom). For the $SIR \times NF$ model, we observe that changing the likelihood that those who recovered from a symptomatic infection become newly fearful of the disease, $\rho_f$, has a dramatic impact on whether or not a second wave occurs. With $\rho_f = 1$ the proportion of the population which are susceptible or fearful largely plateau around 80\% and 60\%, respectively, after 360 days, as shown in  Fig.~\ref{fig:sir_rho_1}. In contrast, these proportions plateau at around 35\% and 0\%, respectively, when we use $\rho_f = 0$, as shown in Fig.~\ref{fig:sir_rho_0}. In additional, a second wave only occurs in the latter case, peaking a bit after 150 days.

When we introduce asymptomatic disease states, as in the $SI_sI_aR_sR_a \times NF$ model, the situation changes further. Keeping all parameters used in the $\rho_f = 1$ run the same with this model results in 80\% of the population still susceptible and 40\% fearful after 360 days, and no second epidemic wave, as shown in Fig.~\ref{fig:si_ai_sr_sr_a_sigma_f_25}. Changing the susceptibility of fearful agents, $\sigma_f$, from 25\% to 35\%, however, is sufficient to produce a second wave in this model, albeit a smaller one which peaks much later at around 275 days, as shown in Fig.~\ref{fig:si_ai_sr_sr_a_sigma_f_35}. Under this scenario, about half of the population is still susceptible after 360 days, although this proportion is still decreasing. The fearful proportion, while approaching zero by 360 days, does so about 160 days later than in the two wave scenario without asymptomatic infections (Fig.~\ref{fig:sir_rho_0}).

Introducing an exposed state and a pre-symptomatic infectious state into the model produces the $SEPI_sI_aR_sR_a \times NF$ model, and further changes in the solution space. Using the same pair of parameter values as with the $SI_sI_aR_sR_a \times NF$ model results in faster disease spread and decay in fear levels. Fig.~\ref{fig:sepi_ai_sr_sr_a_sigma_f_25} shows that about 70\% and 20\% of the population is susceptible and fearful, respectively, after 360 days with $\sigma_f = 0.25$ in this model. Additionally, while we do not see a second wave peak within 360 days here, we do see the start of such a wave and the lack of a plateau of fear levels that corresponds to a single-wave outbreak in the other models. When using $\sigma_f = 0.35$ with the $SEPI_sI_aR_sR_a \times NF$ model, Fig.~\ref{fig:sepi_ai_sr_sr_a_sigma_f_35} shows that fear levels reach zero and a second epidemic peak occurs, as with the $SI_sI_aR_sR_a \times NF$ model. However, both these events happen much sooner in the former model, although still about 100 days later than in the $SIR \times NF$ model.

Altogether, with respect to \ref{rq:compartments} we find that adding asymptomatic states to the ODE models significantly delays the progression of multiple waves and reduces the rate of infections and fear decay as well as the total attack rate, while adding an exposed state has the opposite effect. Neither addition significantly changes the peak of fear levels. As for \ref{rq:params}, we find that changing the values of $\rho_f$ and $\sigma_f$ both showed the potential to change both whether or not we observe a second wave and the total attack rate.

\subsection{Comparing selected scenarios in EpiCast}
\label{sec:scenarios}

\begin{figure*}[t]
    \centering
    \begin{subfigure}{.45\textwidth}
        \centering
        \includegraphics[width=\textwidth]{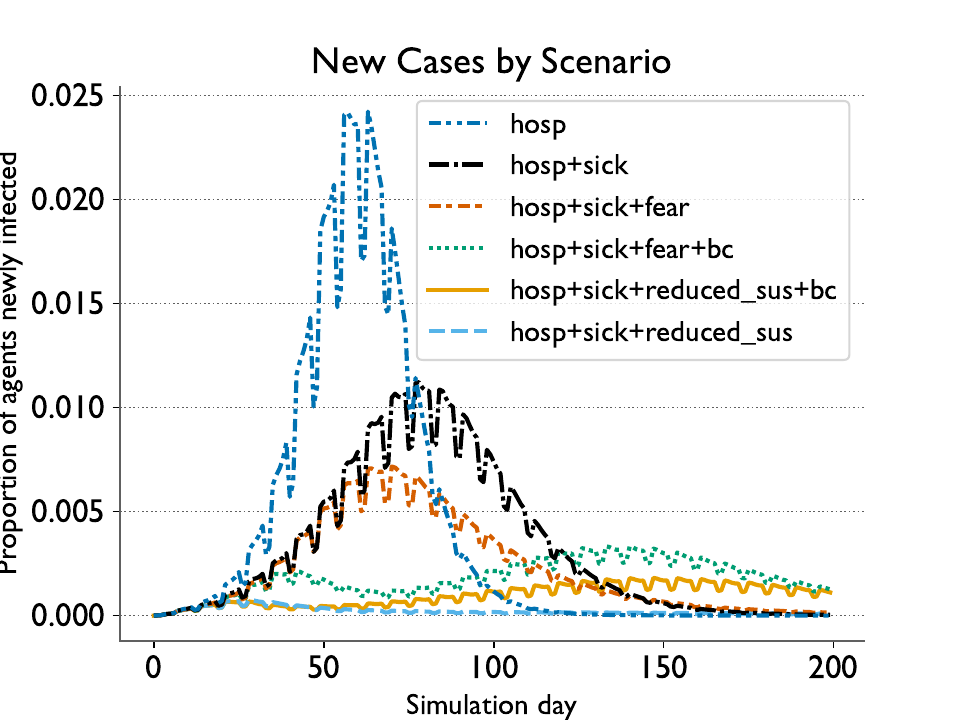}
        \caption{}
        \label{fig:cases}
    \end{subfigure}
    \begin{subfigure}{.45\textwidth}
        \centering
        \includegraphics[width=\textwidth]{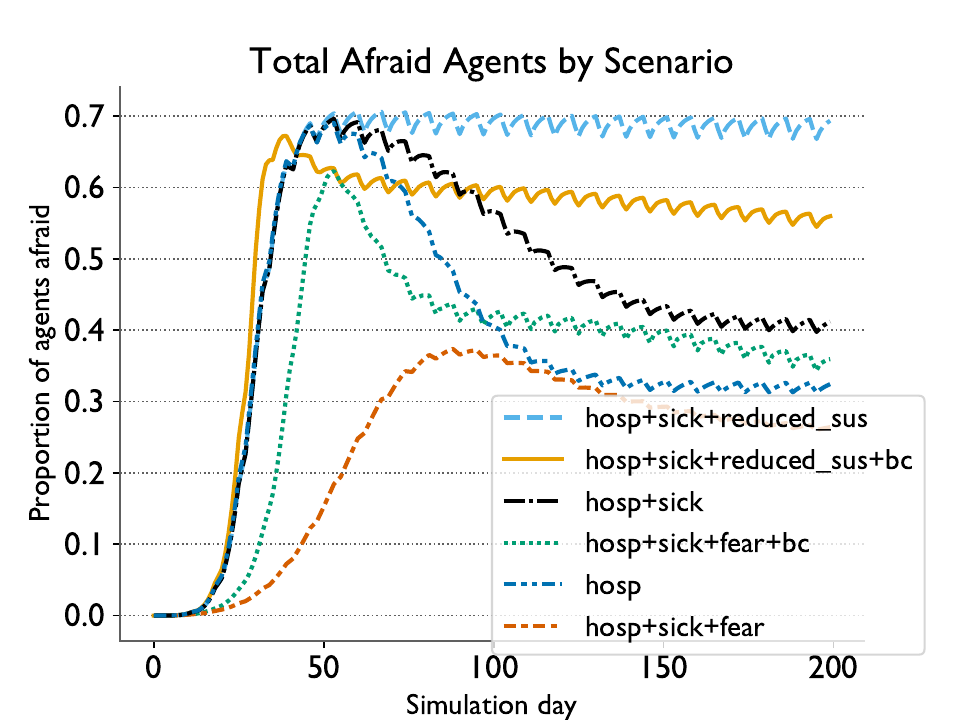}
        \caption{}
        \label{fig:fear}
    \end{subfigure}
    \\
    \begin{subfigure}{.45\textwidth}
        \centering
        \includegraphics[width=\textwidth]{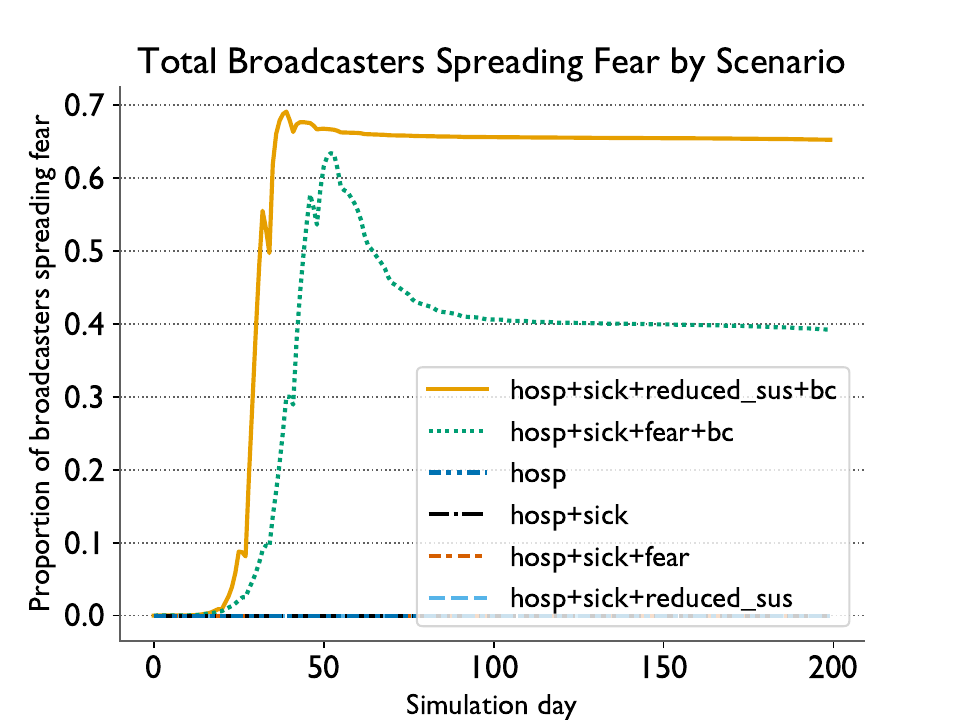}
        \caption{}
        \label{fig:bc-spread}
    \end{subfigure}
    \begin{subfigure}{.45\textwidth}
        \centering
        \includegraphics[width=\textwidth]{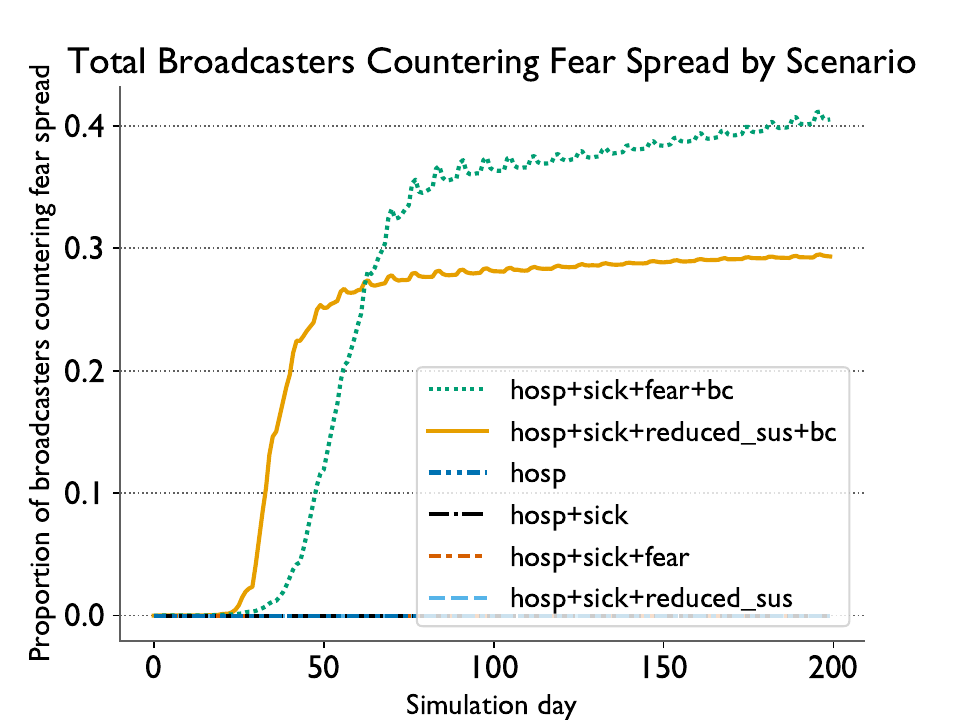}
        \caption{}
        \label{fig:bc-counter}
    \end{subfigure}

    \caption{{\bf EpiCast Fear Spread Scenarios:} New cases (\ref{fig:cases}), total fear levels (\ref{fig:fear}), total broadcasters spreading fear (\ref{fig:bc-spread}), and total broadcasters countering fear spread (\ref{fig:bc-counter}) for six different scenarios run in EpiCast. These scenarios can include agents withdrawing from their normal schedules due to being hospitalized (hosp), being fearful of the disease while having symptoms of it (sick), and fear of the disease without symptoms (fear), as well as non-local fear spread through broadcast media (bc), and fearful agents having a lower susceptibility to the disease due to taking protective actions such as masking (reduced\_sus).}
    \label{fig:scenarios}
\end{figure*}

We now examine the results of the various scenarios evaluated for the coupled contagion model implemented in EpiCast, so as to explore \ref{rq:scenarios}. Fig.~\ref{fig:cases} showcases the differences in the count, height, and timing of epidemic peaks between scenarios. In particular, we see the highest peak in the scenario with only hospitalization-based withdrawals (hosp), as well as the highest total attack rate. This is followed by the scenario where we allow withdrawals based on having symptoms (hosp+sick) and pure-fear withdrawals from fear as well (hosp+sick+fear), which also show a single peak. The scenario with the smallest peak (and also the lowest total attack rate) comes from using hospitalization-based and symptomatic withdrawals along with a susceptibility reduction from fear (hosp+sick+reduced\_sus), which also has a single peak. The only scenarios wherein we observe multiple peaks are those where we add broadcasters as a fear spread mechanism with purely fear-based withdrawals (hosp+sick+fear+bc)and a fear-based susceptibility reduction (hosp+sick+reduced\_sus+bc), with the former having higher peaks than the latter, and the second peak occurring at roughly the same time in both cases.

We observe that fear levels peak after 35 to 60 days at around 60-70\% of the population in most cases, with the exception of the pure-fear withdrawal scenario without broadcasters (hosp+sick+fear), where fear levels peak after 85 days at around 38\% of the population. In most cases, fear levels fall dramatically from this peak before the end of the simulation. The scenarios with reduced susceptibility due to fear both buck this trend, with fear levels only falling slightly by day 200 without broadcasters (hosp+sick+reduced\_sus) and dropping slowly by around 15 percentage points in total with broadcasters (hosp+sick+reduced\_sus+bc). We note that, in the absence of broadcasters, the reduced susceptibility case (hosp+sick+reduced\_sus in Fig.~\ref{fig:fear}) displays the highest fear levels for most of the simulation and the pure-fear withdrawal case the lowest (hosp+sick+fear in Fig.~\ref{fig:fear}). In the presence of broadcasters, fear trends more closely resemble other runs. Fear levels fall significantly after their peak in the susceptibility reduction case (hosp+sick+reduced\_sus+bc in Fig.~\ref{fig:fear}), when they remaining roughly constant in the absence of broadcasters. Conversely, fear levels rose much higher before falling in the pure-fear withdrawal case (hosp+sick+fear+bc in Fig.~\ref{fig:fear}), peaking at above 60\% of the population rather than less than 40\%.

In both scenarios with broadcasters, more broadcasters spread fear than countered it throughout the majority of the run, though in the pure-fear withdrawal case this changed shortly before the end of the run. Far more broadcasters were spreading fear in the susceptibility reduction scenario than in the pure-fear withdrawal scenario for the entire duration (compare hosp+sick+reduced\_sus+bc and hosp+sick+fear+bc in Fig.~\ref{fig:bc-spread}), while the reverse is true for fear countering for most of the duration of the run (Fig.~\ref{fig:bc-counter}). Broadcasters were not used in the other scenarios.

\begin{figure*}[t]
    \centering
    \begin{subfigure}{0.59\textwidth}
      \centering
      \includegraphics[height=2.25in]{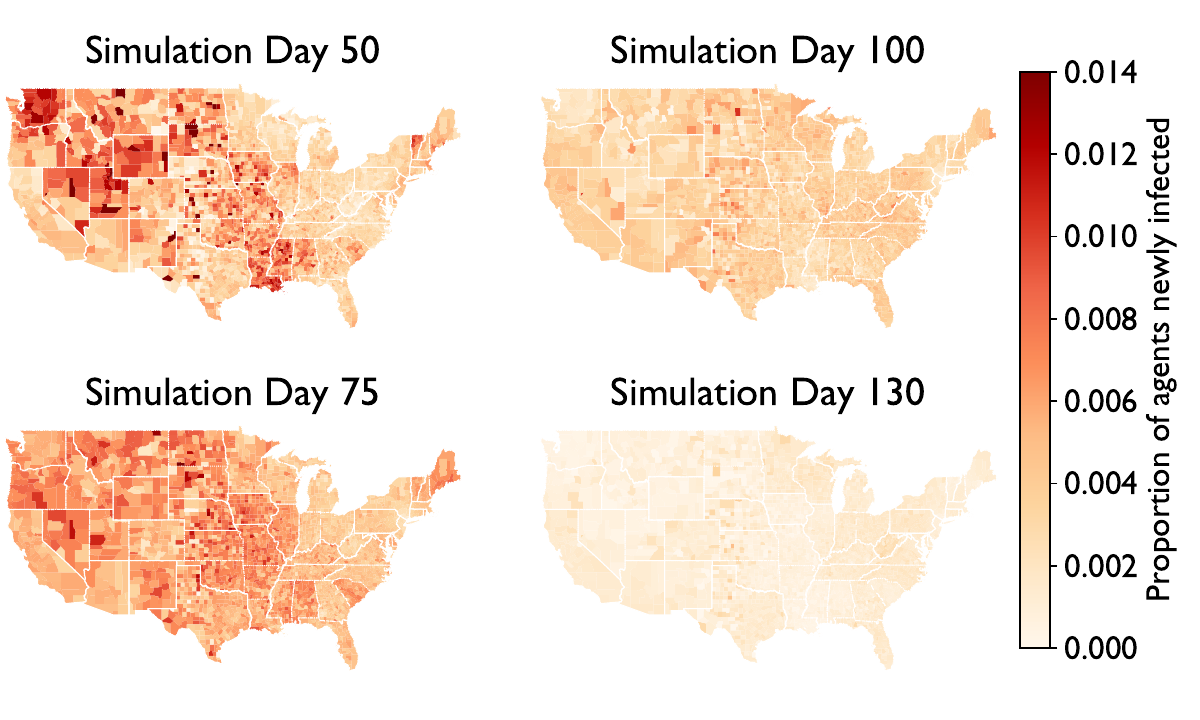}
      \caption{}
      \label{fig:geo-map-l}
    \end{subfigure}
    \begin{subfigure}{0.4\textwidth}
      \centering
      \includegraphics[height=2.25in]{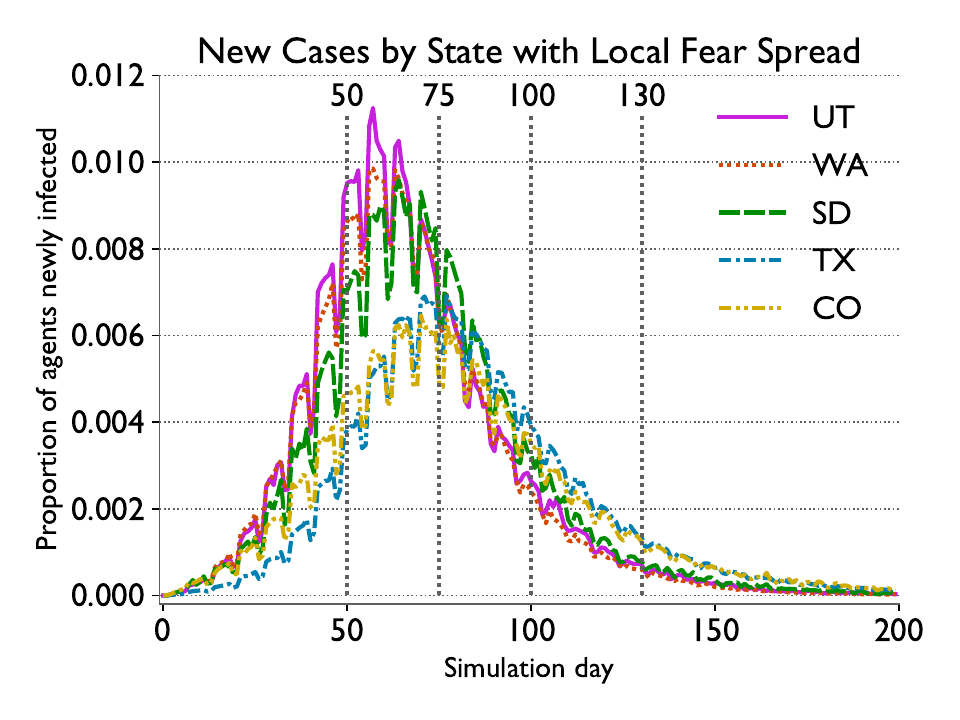}
      \caption{}
      \label{fig:geo-overview-l}
    \end{subfigure}
    \\
    \begin{subfigure}{0.59\textwidth}
      \centering
      \includegraphics[height=2.25in]{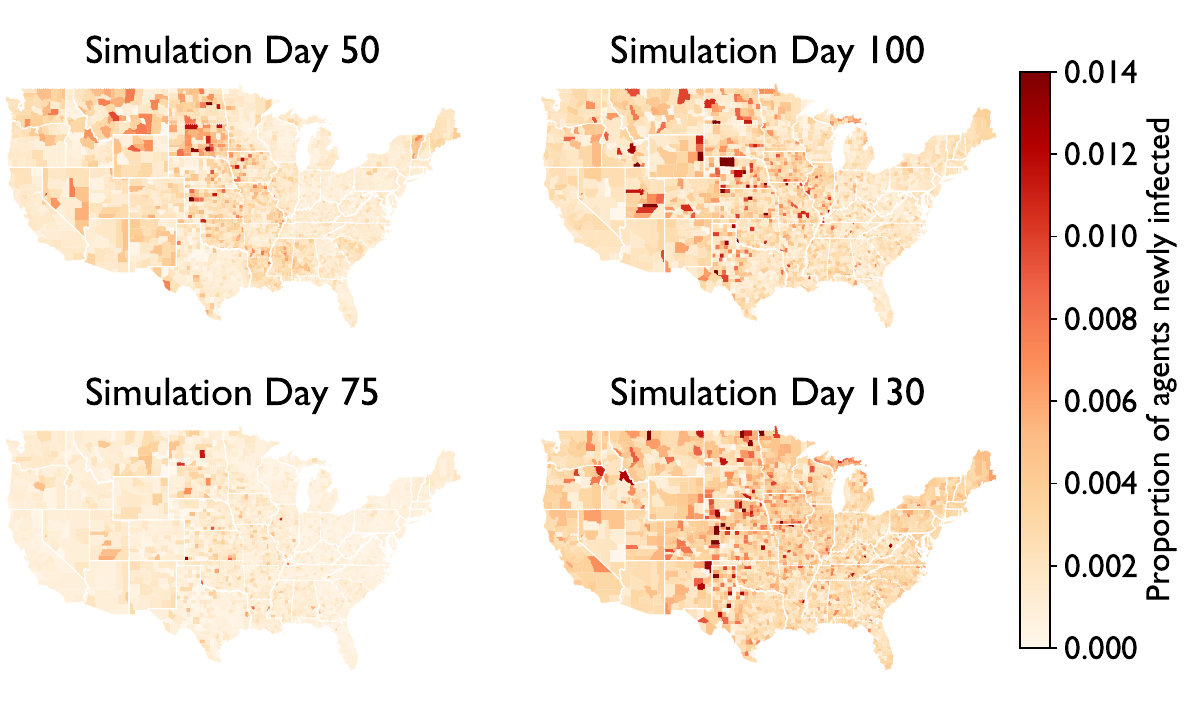}
      \caption{}
      \label{fig:geo-map-bc}
    \end{subfigure}
    \begin{subfigure}{0.4\textwidth}
      \centering
      \includegraphics[height=2.25in]{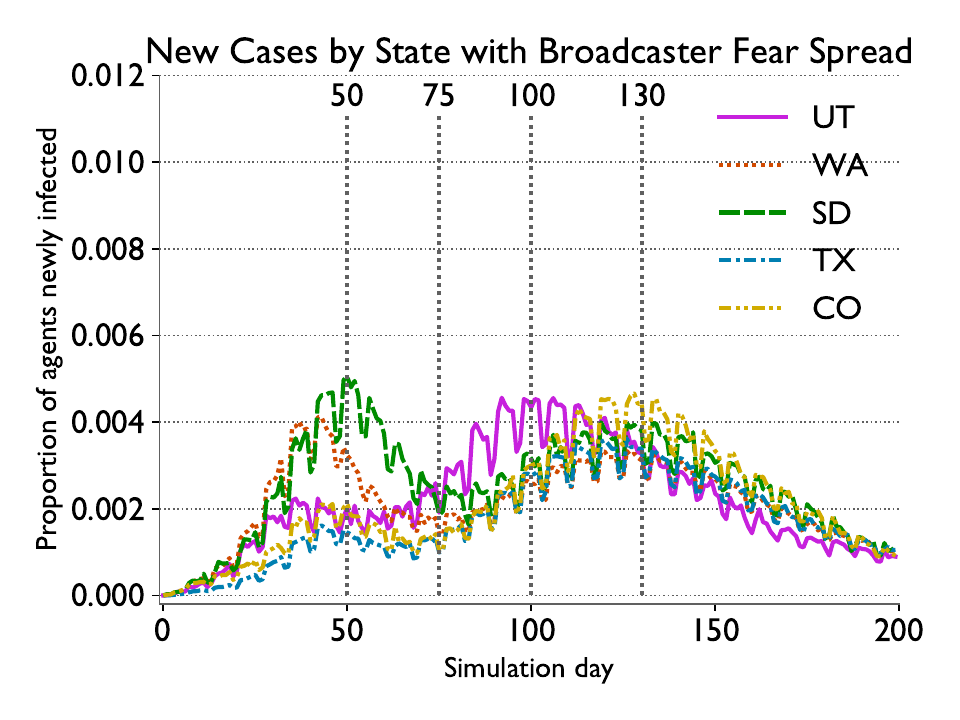}
      \caption{}
      \label{fig:geo-overview-bc}
    \end{subfigure}

    \caption{{\bf Geographic Distribution of Outbreaks in Two Epicast Fear Spread Scenarios:} Geographic distribution of new case counts for selected timestamps (left) and overall new case trends for selected states (right) for scenarios with pure-fear withdrawals without (upper) and with (lower) broadcaster-based fear spread.}
    \label{fig:geo}
\end{figure*}

Fig.~\ref{fig:geo} highlights the geographical distribution of new cases for selected time steps and US states for two of the scenarios shown above. Fig.~\ref{fig:geo-map-l} and \ref{fig:geo-overview-l} represent the results of the scenario with pure-fear withdrawals from fear with purely local fear spread, where Fig.~\ref{fig:geo-map-bc} and \ref{fig:geo-overview-bc} do so when fear spread through broadcasters is additionally allowed (hosp+sick+fear and hosp+sick+fear+bc, respectively, in Fig.~\ref{fig:scenarios}). The time steps chosen represent peaks (days 50, 100, and 130) and a trough (day 75) in the latter, multi-wave scenario. Particular US states were chosen to represent different epidemic trajectories.

For the former case, Fig.~\ref{fig:geo-overview-l} shows how the state-level trajectories differ primarily based on the height of their peak, though Texas (TX) does generally lag behind the trends displayed by the other selected states by a handful of days, with the exception of Colorado (CO) in the declining phase of the outbreak.

In contrast, we see significant divergence between states in the broadcaster scenario, seen in Fig.~\ref{fig:geo-overview-bc}. South Dakota (SD) and Washington (WA) both have high initial peaks which transition into lower secondary peaks, with WA having an earlier first peak and consistently lower case counts throughout. Utah (UT), Colorado (CO), and Texas (TX), by contrast, each have low initial peaks and higher secondary peaks. For UT, there is very little decline in cases before the upward trend towards the second peak begins, with case counts peaking roughly 20-30 days before the other states. UT and CO reach similar heights to each other for each peak, while TX has a lower second peak, only slightly above WA.

\subsection{EpiCast sensitivity analysis}
\label{sec:params}

In order to get a better sense of how varying the behavioral response to fear within EpiCast influences the overall trajectory of disease outbreaks, as per \ref{rq:params}, we ran a number of simulations varying two influential parameters. The first was the likelihood that an agent who is fearful of the disease -- but not currently showing symptoms -- withdraws from their regular schedule of activities purely due to fear, $p_\text{fear}$. The second was the scaling factor applied to an fearful agent's susceptibility to infection, $\sigma_f$, which represents the impact of fearful agents adopting protective behaviors that do not directly change their daily schedule, such as masking, maintaining social distance while in public, and hand-washing.

\begin{figure*}
    \centering
    \begin{subfigure}{0.85\textwidth}
        \centering
        \includegraphics[width=\textwidth]{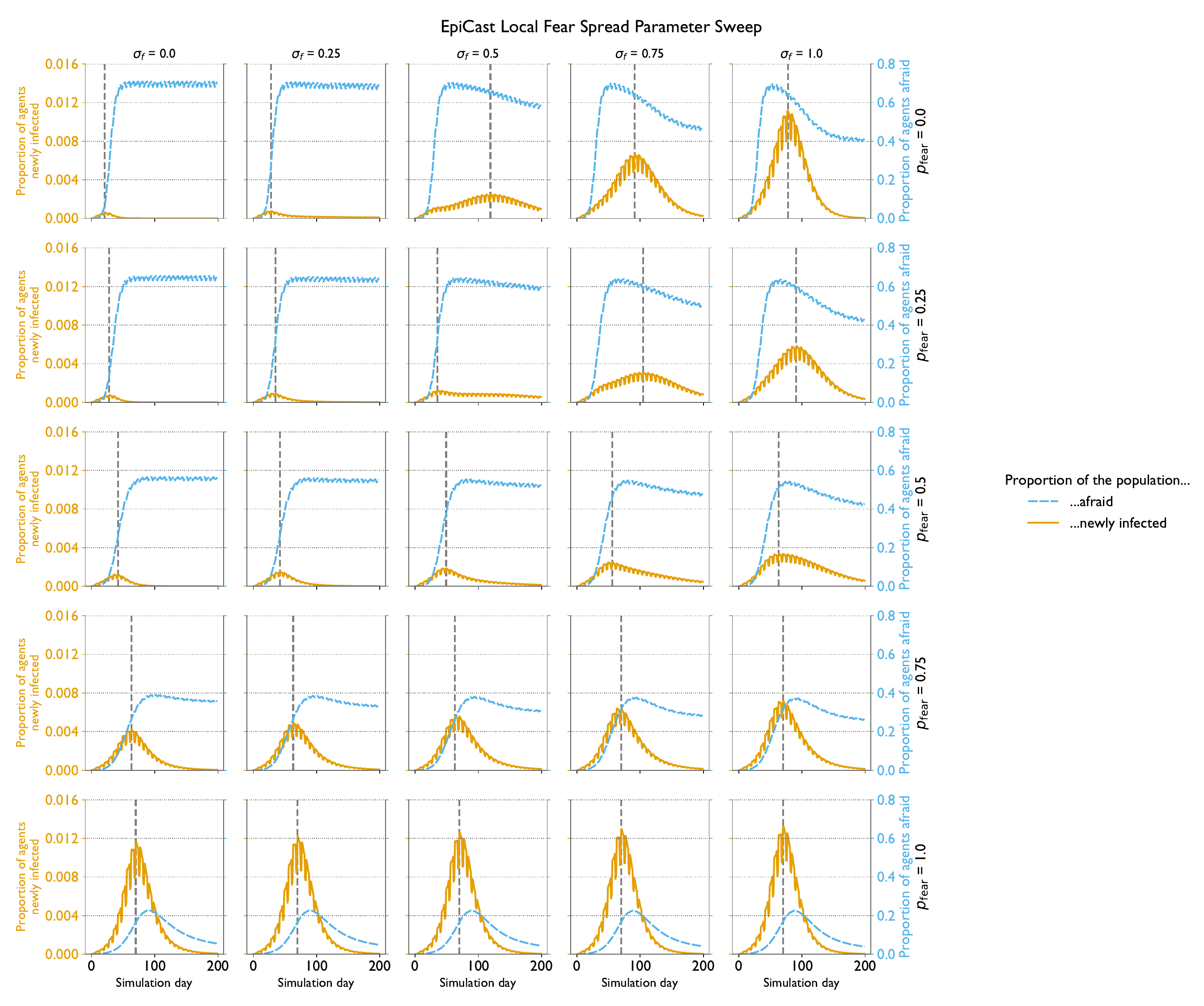}
        \caption{}
        \label{fig:sa-l-grid}
    \end{subfigure}\\
    \begin{subfigure}{0.45\textwidth}
        \centering
        \includegraphics[width=.88\textwidth]{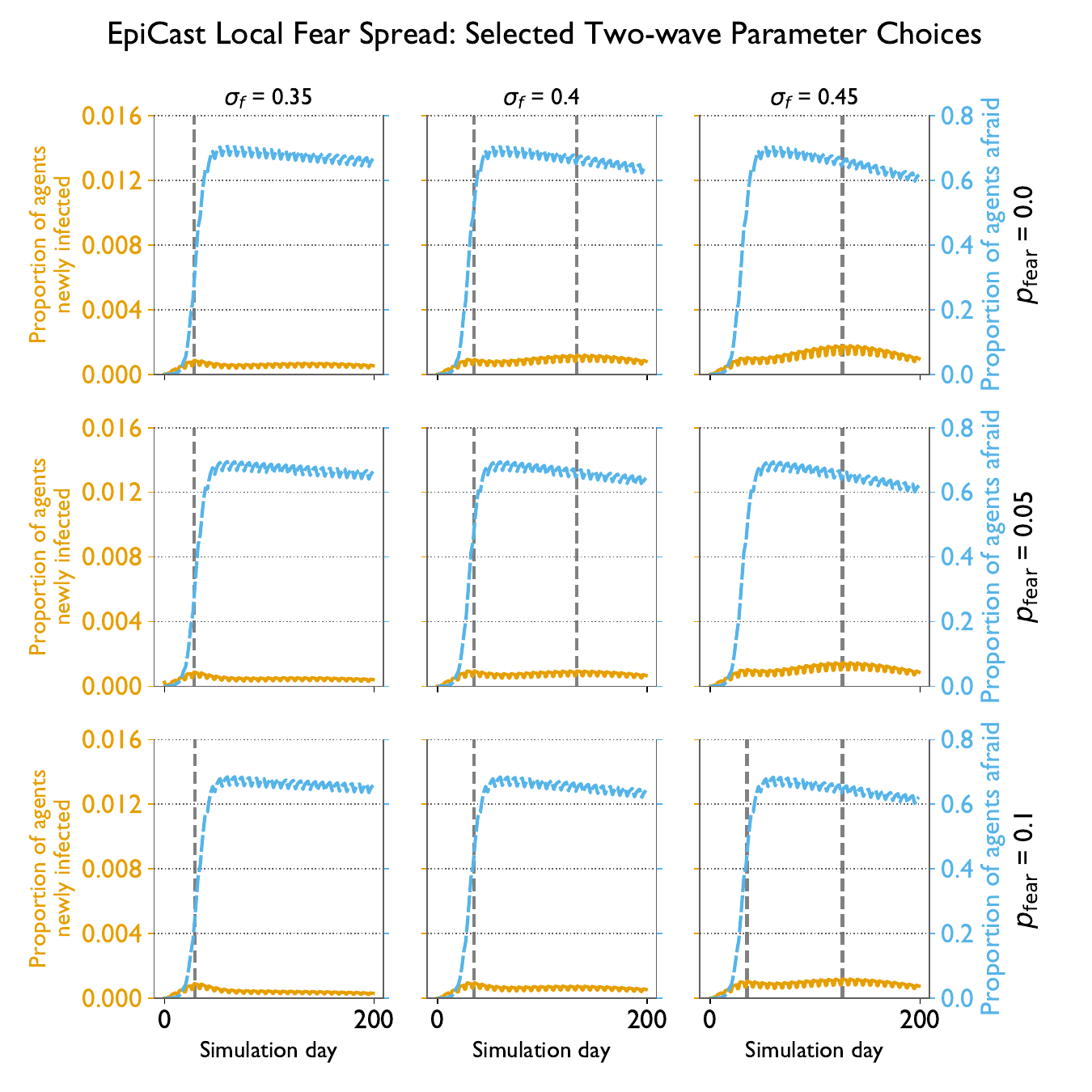}
        \caption{}
        \label{fig:sa-l-sup}
    \end{subfigure}
    \begin{subfigure}{0.45\textwidth}
        \centering
        \includegraphics[width=\textwidth]{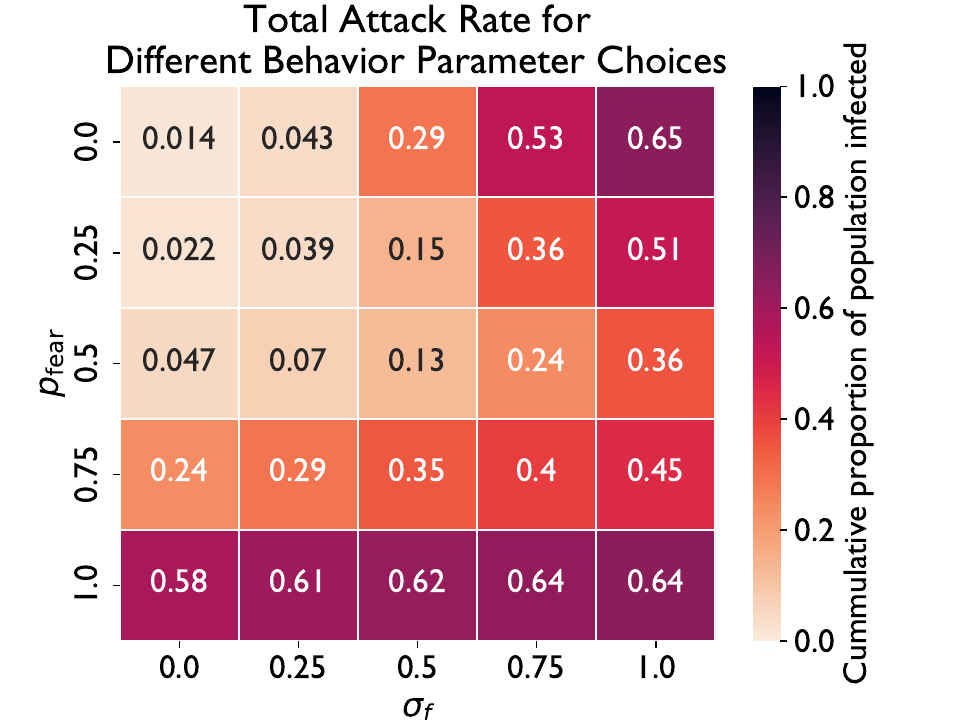}
        \caption{}
        \label{fig:sa-l-attack}
    \end{subfigure}

    \caption{{\bf Epicast local fear spread sensitivity analysis:} New cases by day (with peaks indicated by dashed vertical line) for (\ref{fig:sa-l-grid}) a coarse-grained parameter search and (\ref{fig:sa-l-sup}) a finer-grained search highlighting cases with multiple waves, and (\ref{fig:sa-l-attack}) total attack rate for various rates of purely fear-based withdrawals ($p_\text{fear}$) and levels of susceptibility scaling due to fear ($\sigma_f$) with only local fear spread. \textcolor{revisions}{This parameter search compares different levels of susceptibility to infection for fearful agents ($\sigma_f$) and rates of withdrawal by fearful agents ($p_\text{fear}$).}}
    \label{fig:sa-l}
\end{figure*}

The results of conducting this parameter sweep with only local fear spread are shown in Fig.~\ref{fig:sa-l}. Fig.~\ref{fig:sa-l-grid} displays the overall trajectories of each outbreak, with the proportion of new cases shown in orange, that of fearful agents show in blue, and epidemic peaks indicated by the dashed vertical lines. Note that all parameter combinations produce a single epidemic wave. The simulation which comes the closest corresponds to $p_\text{fear} = 0$ and $\sigma_f = 0.5$. Further experiments, shown in Fig.~\ref{fig:sa-l-sup}, revealed that a very narrow range of values -- $p_\text{fear} = 0, 0.05, \sigma_f = 0.4$, and $p_\text{fear} = 0.1, \sigma_f = 0.45$ -- produce multiple waves, but that a deviation by as little $0.05$ in either parameter can be sufficient to return to the single wave scheme.
Fig.~\ref{fig:sa-l-attack} shows the total attack rate of each simulated outbreak.

We observe the highest attack rate with $p_\text{fear}=0$ and $\sigma_f = 1$, which represents a minimal behavioral response to fear (and corresponds to Scenario (b) from Section~\ref{sec:scenarios}), followed by the cases with $p_\text{fear} = 1$ with lower values of $\sigma_f$. Notably, fear levels are much lower in the latter case. The lowest attack rates, in contrast, are observed when $p_\text{fear}$ and $\sigma_f$ are both small.

Higher values of $p_\text{fear}$ correspond to fewer fearful agents and result reduced variation in fear trajectories for different values of $\sigma_f$. For $p_\text{fear} \leq 0.5$, we observe that fear levels rapidly plateau for $\sigma_f \leq 0.25$ and rise to a peak before falling to varying degrees for $\sigma_f \geq 0.5$. This drop in fear levels is more pronounced for larger values of $\sigma_f$, when the overall attack rate is higher.

\begin{figure*}
    \centering
    \begin{subfigure}{0.85\textwidth}
        \centering
        \includegraphics[width=\textwidth]{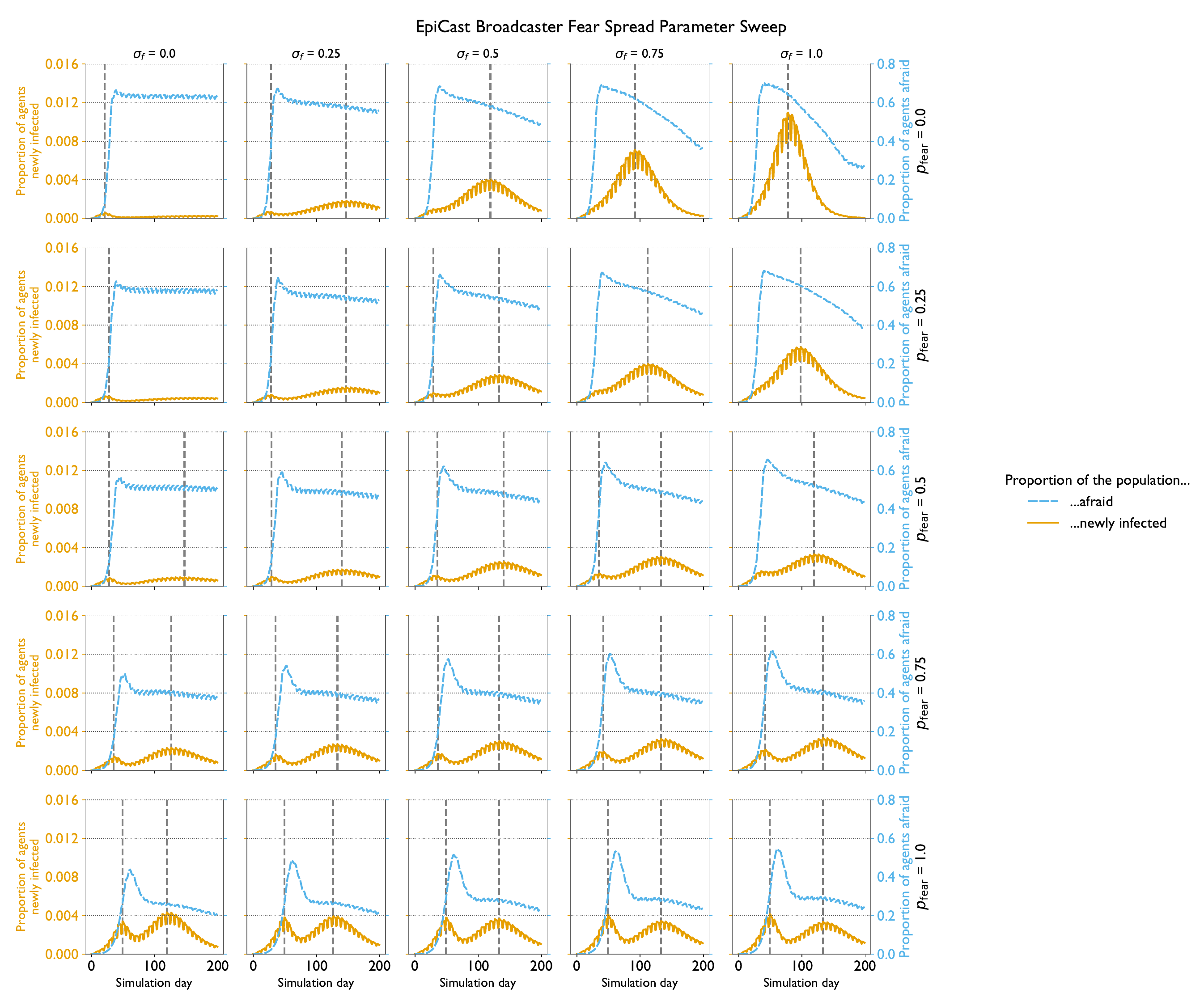}
        \caption{}
        \label{fig:sa-lb-grid}
    \end{subfigure}\\
    \begin{subfigure}{0.45\textwidth}
        \centering
        \includegraphics[width=\textwidth]{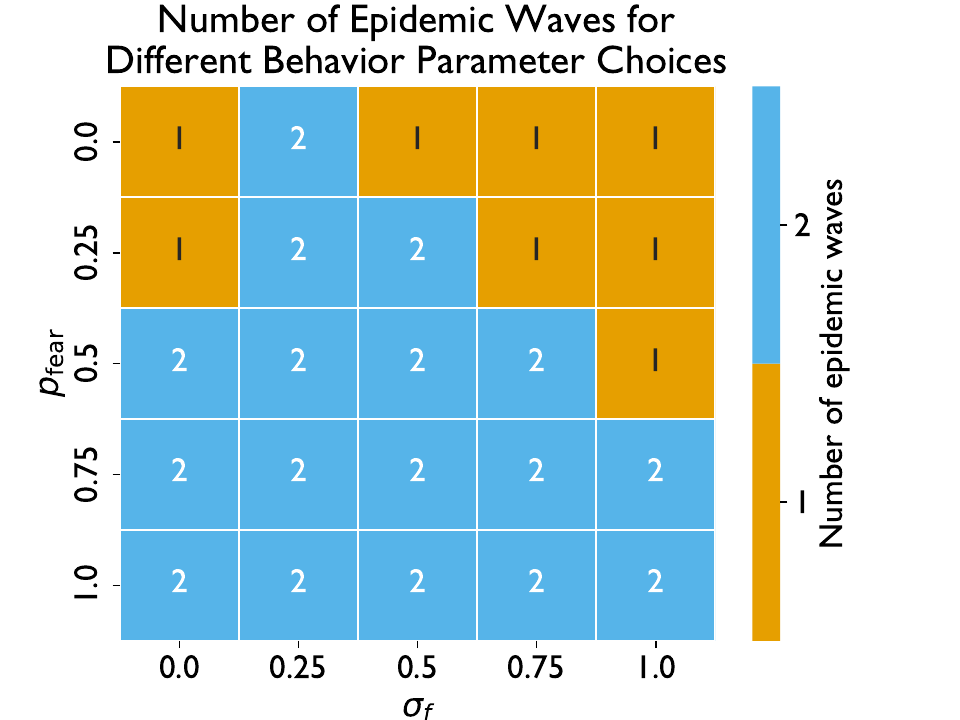}
        \caption{}
        \label{fig:sa-lb-waves}
    \end{subfigure}
    \begin{subfigure}{0.45\textwidth}
        \centering
        \includegraphics[width=\textwidth]{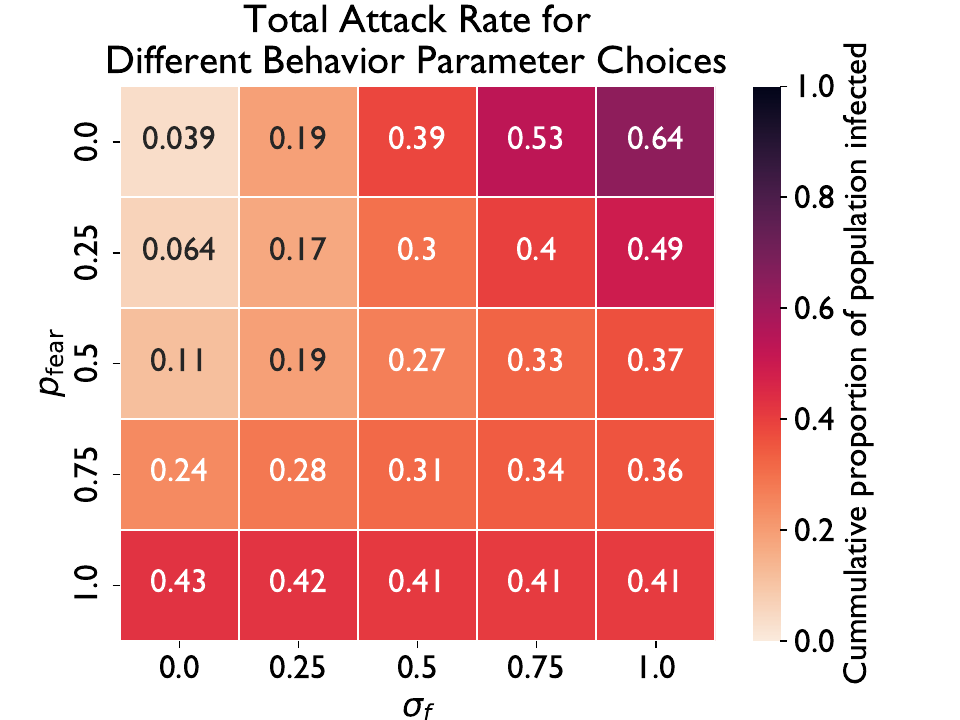}
        \caption{}
        \label{fig:sa-lb-attack}
    \end{subfigure}

    \caption{{\bf Epicast broadcaster fear spread sensitivity analysis:} (\ref{fig:sa-lb-grid}) new cases by day (with peaks indicated by dashed vertical line), (\ref{fig:sa-lb-waves}) number of epidemic waves, and (\ref{fig:sa-lb-attack}) total attack rate for various rates of purely fear-based withdrawals ($p_\text{fear}$) and levels of susceptibility scaling due to fear ($\sigma_f$) with both local and broadcaster-based fear spread. \textcolor{revisions}{This parameter search compares different levels of susceptibility to infection for fearful agents ($\sigma_f$) and rates of withdrawal by fearful agents ($p_\text{fear}$).}}
    \label{fig:sa-lb}
\end{figure*}

The results of conducting this parameter sweep with both local and broadcaster-based fear spread are shown in Fig.~\ref{fig:sa-lb}. The overall trajectories of each outbreak are shown in Fig.~\ref{fig:sa-lb-grid} with the dashed vertical lines indicating where peaks occur, and the orange and blue lines indicating proportion of new cases and fearful agents, respectively. Fig.~\ref{fig:sa-lb-waves} highlights which combinations of parameters produce one and two epidemic waves. All single wave scenarios occur with low values of $p_\text{fear}$, and either a high or low value of $\sigma_f$ (i.e the upper right corner of Fig.~\ref{fig:sa-lb-waves} and the top two cells in the left of that figure). All other parameter combinations produce two waves. These two waves are generally about the same height with $p_\text{fear} = 1$ but the second is both taller and longer than the first in most other cases.

The same parameter combinations that produce a single wave also produce the most extreme attack rates, as highlighted by Fig.~\ref{fig:sa-lb-attack}. The upper right corner (with high values of $\sigma_f$ and low values of $p_\text{fear}$) contains cells with the highest total attack rates, while the single wave runs in the upper left corner exhibit the lowest. Compared to the corresponding simulations without broadcaster-based fear spread, those with broadcaster-spread fear tend to show more moderate attack rates. For example, attack rates fall from $58\%-64\%$ (bottom row of Fig.~\ref{fig:sa-l-attack}) to $43\%-41\%$ (bottom row of Fig.~\ref{fig:sa-lb-attack}) when broadcasters are introduced with $p_\text{fear} = 1.0$. Conversely, attack rates largely rise in the top half of the first two columns the figure, with for instance the attack rate with $p_\text{fear} = 0, \sigma_f = 0$ increasing from 1.4\% to 3.9\%, and that with $p_\text{fear} = 0.25, \sigma_f = 0.25$ rising from $7\% to 19\%$.

Fear levels, in turn, generally follow the same basic trajectory, rising to an initial peak before falling in first  an initial steep decline, then a secondary slow decline. The transition between these two periods of decline generally corresponds to when cases begin rising for a second epidemic wave. The main exceptions are found in the top left and right corners of Fig.~\ref{fig:sa-lb-grid}, a plateau after a very short  decline from the initial peak in the former case and a single accelerating decline for two cells in the latter. These mostly correspond to the cells with a single epidemic wave. The height of the initial peaks decreases as $p_\text{fear}$ increases and fear levels fall farther when more cases occur.

\section{Discussion}

Examining the fear levels in each scenario, shown in Fig.~\ref{fig:fear}, offers some insights into why we might observe multiple peaks primarily in cases with non-local fear spread via broadcasters. In cases where fear reduces agent susceptibility to infection, in the absence of fear spread through broadcasters (hosp+sick+reduced\_sus) fear levels never fall far below their peak, at around 50 days. This is likely due to infection levels never rising to the point where there is a sufficient number of agents in the $R_s$ state to cause fearful agents to lose their fear. This is in contrast to what occurs with fear spread through broadcasters (hosp+sick+reduced\_sus+bc), where fear levels quickly fall after reaching their peak. This suggests that the influence of broadcasters is sufficient to lower fear levels to the point where the disease can start spreading again, causing a second wave. Roughly the opposite appears to occur in the cases where we allow purely fear-based withdrawals; without the influence of broadcasters (hosp+sick+fear), fear levels remain lower than in any other scenario for the duration of the run. This is likely because withdrawals reduce \emph{both} fear and disease spread in this mode and there is a high chance that any given fearful person will withdraw. Conversely, with the influence of broadcasters (hosp+sick+fear+bc) fear levels rise much higher before falling, though they remain lower than in all other scenarios except that with only hospitalization-based withdrawals. This suggests broadcasters allow fear to spread even when most fearful agents remain withdrawn.

These explanations are supported by examining the number of broadcasters spreading (see Fig.~\ref{fig:bc-spread}) and countering fear spread (see Fig.~\ref{fig:bc-counter}) over time. The period from days 25-50 where fear levels peak and then begin to fall in the reduced susceptibility scenario with broadcasters is also the period when we see the number of broadcasters countering fear spread dramatically increase, before reaching a much slower sustained growth rate. This slower growth period corresponds to a slow linear decrease in the number of fearful agents. The number of fear spreading broadcasters follows a similar trend to fear levels in this case, albeit with a smaller falloff after the peak. Similarly, we see that overall fear levels and the number of fear-spreading broadcasters show similar patterns in the pure-fear withdrawal case, with the dramatic reduction in fearful agents between days 50 and 100 matched by decline in fear-spreading broadcasters of similar magnitude.

In addition, the very narrow range of parameter values for which we do observe multiple waves with purely local fear spread appears to fall in the midst of a phase transition between a regime of small, short outbreaks and large, long outbreaks as represented by the cell $p_\text{fear} = 0, \sigma_f = 0.25$, and the cell $p_\text{fear} = 0, \sigma_f = 0.75$ in Fig.~\ref{fig:sa-l-grid}, respectively. The former regime corresponds to lower values of $\sigma_f$ which represent a strong NPI response to fear, implying that fear-induced behavior reduces disease spread sufficiently to forestall a larger outbreak here. The latter regime, by contrast, represents minimal behavioral responses to fear, as seen by the fact it centers on $p_\text{fear} = 0, \sigma_f = 1$, where fear has no impact of the behavior of agents without symptoms. Intriguingly, these two regimes also appear with broadcaster-based fear spread, and correspond to the only cells that do not exhibit multiple waves in Fig.~\ref{fig:sa-lb-grid}.

Altogether, these observations suggest that we observe two waves for opposite reasons in these cases. With reduced susceptibility, fear levels generally do not fall far enough for disease transmission to pick back up and produce a second wave without broadcasters countering fear spread. Conversely, with pure-fear fear-based withdrawals, fear levels generally fail to rise high enough fast enough for the first wave to leave a large reservoir of susceptible agents without broadcasters increasing the rate of fear spread. In either case, non-local fear spread, as represented by broadcasters, appears to facilitate a transition from high initial fear levels to low final fear levels in the presence of fear-based behaviors that significantly reduce disease spread. We posit that this transition only occurs under a few narrow range of conditions in the case of purely local fear spread as a result of fear both spreading and receding relatively slowly. In contrast, the faster rate at which fear levels rise and fall in the presence of broadcasters may explain why multiple epidemic waves are more commonly observed under these conditions. In other words, in order to produce multiple epidemic waves in this model, fear must first spread significantly faster than the disease and then recede faster than it as well. Within this framing, it is clear why broadcasters facilitate the formation of multiple waves: information and the behavioral changes it engenders both spreads and recedes more rapidly when it is less tightly constrained by physical distance.

\textcolor{revisions}{These results demonstrate that bottom-up drivers of behavior can shift outcomes in infectious disease simulations even in the absence of top-down interventions, like lockdown orders. Additionally, the difference in dynamics produced by adding non-local fear spread to our model further suggests that different structures of interactions between private actors may produce different behaviors, even when the underlying attitudes of the population remain unchanged. These findings complement observational work from the COVID-19 pandemic. One such study by Pan et al.~\cite{pan2020quantifying} found that social distancing levels began rising prior to lockdown orders being issued and falling before those orders were lifted in many U.S. states. Together, these results highlight the importance of securing broad buy-in from the public when implementing population-wide policy interventions in response to a pandemic.}

The results we present in this paper are perhaps best understood as an exploration of the high-level implications of our model. Though the foundational elements of EpiCast are mature and have been calibrated extensively, there remain significant obstacles to doing so for the fear-based components of the model. While some survey data are available describing various behaviors throughout part of the pandemic~\cite{delphi2022ctis}, it is much more difficult to tie behavior to specific influences or rationales. We thus confine ourselves to presenting the model itself and exploring the impact of varying parameter values, and leave calibration efforts to future work.

In addition, we acknowledge that the parameter sweep based sensitivity analysis study presented above is necessarily incomplete. While we explore the full possible \emph{range} of parameter values for $p_\text{fear}$ and $\sigma_f$, we do so at a relatively coarse resolution, so as to limit the total number of simulation runs. As evidenced by the multi-wave scenarios presented in Fig.~\ref{fig:sa-l-sup}, this coarseness means that this study may miss dynamics that only emerge for a small range of parameter values. In addition, there are many simulation parameters beyond the two we focus on in the sensitivity analysis above. While we explore the impacts of the initial conditions, model stochasticity, and threshold for broadcaster opinion adoption in Supplementary Section 3, extending the parameter sweep method to the full range of model parameters would be prohibitively computationally expensive. In the absence of such an experiment it is unclear how varying those parameters might affect the dynamics in EpiCast. We leave a more detailed exploration of the impact of these remaining parameters to future work.

\section{Conclusion}
In this work we present a series of models which capture the coupled spread of fear of a disease and the disease itself. We do so with several ODE models in order to demonstrate the impact of adding additional disease states to the model, representing asymptomatic, exposed, and pre-symptomatic disease states, and an ABM covering all disease states with both local and non-local mechanisms for fear spread implemented.

We find that the total attack rate is reduced by the addition of asymptomatic states and increased by the addition of a pre-symptomatic state in our ODE models, and the reverse is true for the rate at which an outbreak progresses.

In our ABM implementation in EpiCast, we primarily observe multiple epidemic waves when using our non-local fear spread mechanism, which represents broadcast media (regional information communication, including print, radio, and TV). We find that a combination of (1) high initial fear levels, (2) a strong behavioral response to fear, and (3) low fear levels later in the simulation produce multiple waves in our model. Thus, one potential explanation for why we only observe multiple waves with purely local fear spread for a very small range of parameter choices is that fear generally spreads (and recedes) much more slowly in the absence of broadcasters in our model.

\section*{Acknowledgments}
Many thanks to Nidhi Parikh, Micaela Richter, Jeffery S. Keithley, and Thomas J. Harris for helpful discussions over the course of this work.

\section*{Funding declaration}
This work was supported by the U.S. Department of Energy,
Office of Science, Office of Advanced Scientific Computing
Research and by cooperative agreement CDC-RFA-FT-23-0069 from
the CDC's Center for Forecasting and Outbreak Analytics.
Its contents are solely the responsibility of the authors and
do not necessarily represent the official views of the Centers
for Disease Control and Prevention.

This work was performed at
Los Alamos National Laboratory (LANL), an equal opportunity
employer, which is operated by Triad National Security, LLC,
for the National Nuclear Security Administration
(NNSA) of the U.S. Department of Energy (DOE) under contract
\#19FED1916814CKC. The funders had no role in study design,
data collection and analysis, decision to publish, or preparation
of the manuscript. The findings and conclusions in this report
are those of the authors and do not necessarily represent
the official position of LANL.
This research used resources provided by the Darwin testbed at
LANL which is funded by the Computational Systems and Software
Environments subprogram of LANL's Advanced Simulation and
Computing program (NNSA/DOE).

This material is based upon work supported by the U.S. Department
of Energy, Office of Science, Office of Advanced Scientific
Computing Research, Department of Energy Computational Science
Graduate Fellowship under Award Number DE-SC0021.

\section*{Author contributions statement}

J.K., P.C.A, D.J.B., C.B., S.Y.D.V., and T.C.G. developed the conceptual framework and J.K. wrote the majority of the code for the behavioral models presented in this work. This code included the ODE models, and the extension to EpiCast, with P.C.A. contributing some code to the later. D.J.B. and C.B. provided suggestions regarding the development and evaluation of the ODE models. P.C.A. and T.C.G. provided insights into the original design and implementation of EpiCast which informed the implementation of the behavioral model. J.T. developed and provided access to UrbanPop, the synthetic population on which this study was conducted, and guided that use. J.K., A.B., S.Y.D.V., and T.C.G. conceived the experiments, and J.K. conducted them and analyzed the results. J.K. prepared the initial draft of the manuscript, along with all figures. A.B., S.Y.D.V., and T.C.G. all provided supervision. All authors reviewed and edited the manuscript.

\section*{Data availability statement}
We have publicly released all output data from simulations described in this paper through Zenodo here: \url{https://zenodo.org/uploads/15706577}. The source code for the scripts used to solve the ODE models, analyze all simulation results, and produce the plots shown in this manuscript are publicly available here: \url{https://github.com/lanl/epicast-coupled}. The source code for EpiCast is available from Los Alamos National Laboratory, but restrictions apply to the availability of this code, which was used under license for the current study, and so is not publicly available. The code is, however, available from the corresponding authors upon reasonable request and with permission of Los Alamos National Laboratory.

\section*{Competing interests}
The authors declare no competing interests.

\bibliography{cite}

\begin{thebibliography}{10}
\urlstyle{rm}
\expandafter\ifx\csname url\endcsname\relax
  \def\url#1{\texttt{#1}}\fi
\expandafter\ifx\csname urlprefix\endcsname\relax\def\urlprefix{URL }\fi
\expandafter\ifx\csname doiprefix\endcsname\relax\def\doiprefix{DOI: }\fi
\providecommand{\bibinfo}[2]{#2}
\providecommand{\eprint}[2][]{\url{#2}}

\bibitem{chernozhukov2021causal}
\bibinfo{author}{Chernozhukov, V.}, \bibinfo{author}{Kasahara, H.} \&
  \bibinfo{author}{Schrimpf, P.}
\newblock \bibinfo{journal}{\bibinfo{title}{Causal impact of masks, policies,
  behavior on early covid-19 pandemic in the us}}.
\newblock {\emph{\JournalTitle{Journal of econometrics}}}
  \textbf{\bibinfo{volume}{220}}, \bibinfo{pages}{23--62}
  (\bibinfo{year}{2021}).

\bibitem{navas2022forecasting}
\bibinfo{author}{Navas~Thorakkattle, M.}, \bibinfo{author}{Farhin, S.} \&
  \bibinfo{author}{Khan, A.~A.}
\newblock \bibinfo{journal}{\bibinfo{title}{Forecasting the trends of covid-19
  and causal impact of vaccines using bayesian structural time series and
  arima}}.
\newblock {\emph{\JournalTitle{Annals of Data Science}}}
  \textbf{\bibinfo{volume}{9}}, \bibinfo{pages}{1025--1047}
  (\bibinfo{year}{2022}).

\bibitem{toharudin2021national}
\bibinfo{author}{Toharudin, T.} \emph{et~al.}
\newblock \bibinfo{journal}{\bibinfo{title}{National vaccination and local
  intervention impacts on covid-19 cases}}.
\newblock {\emph{\JournalTitle{Sustainability}}} \textbf{\bibinfo{volume}{13}},
  \bibinfo{pages}{8282} (\bibinfo{year}{2021}).

\bibitem{suthar2022public}
\bibinfo{author}{Suthar, A.~B.} \emph{et~al.}
\newblock \bibinfo{journal}{\bibinfo{title}{Public health impact of covid-19
  vaccines in the us: observational study}}.
\newblock {\emph{\JournalTitle{Bmj}}} \textbf{\bibinfo{volume}{377}}
  (\bibinfo{year}{2022}).

\bibitem{cramer_evaluation_2022}
\bibinfo{author}{Cramer, E.~Y.} \emph{et~al.}
\newblock \bibinfo{journal}{\bibinfo{title}{Evaluation of individual and
  ensemble probabilistic forecasts of {COVID}-19 mortality in the {United}
  {States}}}.
\newblock {\emph{\JournalTitle{Proceedings of the National Academy of
  Sciences}}} \textbf{\bibinfo{volume}{119}}, \bibinfo{pages}{e2113561119},
  \doiprefix\url{10.1073/pnas.2113561119} (\bibinfo{year}{2022}).
\newblock \bibinfo{note}{Publisher: Proceedings of the National Academy of
  Sciences}.

\bibitem{nixon2022evaluation}
\bibinfo{author}{Nixon, K.} \emph{et~al.}
\newblock \bibinfo{journal}{\bibinfo{title}{An evaluation of prospective
  covid-19 modelling studies in the usa: from data to science translation}}.
\newblock {\emph{\JournalTitle{The Lancet Digital Health}}}
  \textbf{\bibinfo{volume}{4}}, \bibinfo{pages}{e738--e747}
  (\bibinfo{year}{2022}).

\bibitem{bauch2013behavioral}
\bibinfo{author}{Bauch, C.}, \bibinfo{author}{d’Onofrio, A.} \&
  \bibinfo{author}{Manfredi, P.}
\newblock \bibinfo{journal}{\bibinfo{title}{Behavioral epidemiology of
  infectious diseases: an overview}}.
\newblock {\emph{\JournalTitle{Modeling the interplay between human behavior
  and the spread of infectious diseases}}} \bibinfo{pages}{1--19}
  (\bibinfo{year}{2013}).

\bibitem{hamilton2024incorporating}
\bibinfo{author}{Hamilton, A.} \emph{et~al.}
\newblock \bibinfo{journal}{\bibinfo{title}{Incorporating endogenous human
  behavior in models of covid-19 transmission: A systematic scoping review}}.
\newblock {\emph{\JournalTitle{Dialogues in Health}}} \bibinfo{pages}{100179}
  (\bibinfo{year}{2024}).

\bibitem{schluter_unraveling_2023}
\bibinfo{author}{Schlüter, M.} \emph{et~al.}
\newblock \bibinfo{journal}{\bibinfo{title}{Unraveling complex causal processes
  that affect sustainability requires more integration between empirical and
  modeling approaches}}.
\newblock {\emph{\JournalTitle{Proceedings of the National Academy of
  Sciences}}} \textbf{\bibinfo{volume}{120}}, \bibinfo{pages}{e2215676120},
  \doiprefix\url{10.1073/pnas.2215676120} (\bibinfo{year}{2023}).
\newblock \bibinfo{note}{Publisher: Proceedings of the National Academy of
  Sciences}.

\bibitem{murakami2022agent}
\bibinfo{author}{Murakami, T.}, \bibinfo{author}{Sakuragi, S.},
  \bibinfo{author}{Deguchi, H.} \& \bibinfo{author}{Nakata, M.}
\newblock \bibinfo{journal}{\bibinfo{title}{Agent-based model using gps
  analysis for infection spread and inhibition mechanism of sars-cov-2 in
  tokyo}}.
\newblock {\emph{\JournalTitle{Scientific Reports}}}
  \textbf{\bibinfo{volume}{12}}, \bibinfo{pages}{20896} (\bibinfo{year}{2022}).

\bibitem{germann2006mitigation}
\bibinfo{author}{Germann, T.~C.}, \bibinfo{author}{Kadau, K.},
  \bibinfo{author}{Longini~Jr, I.~M.} \& \bibinfo{author}{Macken, C.~A.}
\newblock \bibinfo{journal}{\bibinfo{title}{Mitigation strategies for pandemic
  influenza in the united states}}.
\newblock {\emph{\JournalTitle{Proceedings of the National Academy of
  Sciences}}} \textbf{\bibinfo{volume}{103}}, \bibinfo{pages}{5935--5940}
  (\bibinfo{year}{2006}).

\bibitem{kersting2021predicting}
\bibinfo{author}{Kersting, M.}, \bibinfo{author}{Bossert, A.},
  \bibinfo{author}{S{\"o}rensen, L.}, \bibinfo{author}{Wacker, B.} \&
  \bibinfo{author}{Schl{\"u}ter, J.~C.}
\newblock \bibinfo{journal}{\bibinfo{title}{Predicting effectiveness of
  countermeasures during the covid-19 outbreak in south africa using
  agent-based simulation}}.
\newblock {\emph{\JournalTitle{Humanities and Social Sciences Communications}}}
  \textbf{\bibinfo{volume}{8}}, \bibinfo{pages}{1--15} (\bibinfo{year}{2021}).

\bibitem{bosman2024agent}
\bibinfo{author}{Bosman, M.} \emph{et~al.}
\newblock \bibinfo{journal}{\bibinfo{title}{An agent based simulation of
  covid-19 history in catalonia using extensive real datasets}}.
\newblock {\emph{\JournalTitle{Scientific Reports}}}
  \textbf{\bibinfo{volume}{14}}, \bibinfo{pages}{31858} (\bibinfo{year}{2024}).

\bibitem{chen2024role}
\bibinfo{author}{Chen, J.} \emph{et~al.}
\newblock \bibinfo{journal}{\bibinfo{title}{Role of heterogeneity: National
  scale data-driven agent-based modeling for the us covid-19 scenario modeling
  hub}}.
\newblock {\emph{\JournalTitle{Epidemics}}} \textbf{\bibinfo{volume}{48}},
  \bibinfo{pages}{100779} (\bibinfo{year}{2024}).

\bibitem{mao2014modeling}
\bibinfo{author}{Mao, L.}
\newblock \bibinfo{journal}{\bibinfo{title}{Modeling triple-diffusions of
  infectious diseases, information, and preventive behaviors through a
  metropolitan social network—an agent-based simulation}}.
\newblock {\emph{\JournalTitle{Applied Geography}}}
  \textbf{\bibinfo{volume}{50}}, \bibinfo{pages}{31--39}
  (\bibinfo{year}{2014}).

\bibitem{alexanderEpicast20Largescale2025}
\bibinfo{author}{Alexander, P.~C.} \emph{et~al.}
\newblock \bibinfo{title}{Epicast 2.0: {{A}} large-scale, demographically
  detailed, agent-based model for simulating respiratory pathogen spread in the
  {{United States}}}, \doiprefix\url{10.48550/arXiv.2504.03604}
  (\bibinfo{year}{2025}).
\newblock \eprint{2504.03604}.

\bibitem{rahmandad2022enhancing}
\bibinfo{author}{Rahmandad, H.}, \bibinfo{author}{Xu, R.} \&
  \bibinfo{author}{Ghaffarzadegan, N.}
\newblock \bibinfo{journal}{\bibinfo{title}{Enhancing long-term forecasting:
  Learning from covid-19 models}}.
\newblock {\emph{\JournalTitle{PLoS computational biology}}}
  \textbf{\bibinfo{volume}{18}}, \bibinfo{pages}{e1010100}
  (\bibinfo{year}{2022}).

\bibitem{tovissode2024heterogeneous}
\bibinfo{author}{Tovissod{\'e}, C.~F.} \& \bibinfo{author}{Baumgaertner, B.}
\newblock \bibinfo{journal}{\bibinfo{title}{Heterogeneous risk tolerance,
  in-groups, and epidemic waves}}.
\newblock {\emph{\JournalTitle{Frontiers in applied mathematics and
  statistics}}} \textbf{\bibinfo{volume}{10}}, \bibinfo{pages}{1360001}
  (\bibinfo{year}{2024}).

\bibitem{kassa2011epidemiological}
\bibinfo{author}{Kassa, S.~M.} \& \bibinfo{author}{Ouhinou, A.}
\newblock \bibinfo{journal}{\bibinfo{title}{Epidemiological models with
  prevalence dependent endogenous self-protection measure}}.
\newblock {\emph{\JournalTitle{Mathematical biosciences}}}
  \textbf{\bibinfo{volume}{229}}, \bibinfo{pages}{41--49}
  (\bibinfo{year}{2011}).

\bibitem{huang2022game}
\bibinfo{author}{Huang, Y.} \& \bibinfo{author}{Zhu, Q.}
\newblock \bibinfo{journal}{\bibinfo{title}{Game-theoretic frameworks for
  epidemic spreading and human decision-making: A review}}.
\newblock {\emph{\JournalTitle{Dynamic Games and Applications}}}
  \textbf{\bibinfo{volume}{12}}, \bibinfo{pages}{7--48} (\bibinfo{year}{2022}).

\bibitem{epstein2008coupled}
\bibinfo{author}{Epstein, J.~M.}, \bibinfo{author}{Parker, J.},
  \bibinfo{author}{Cummings, D.} \& \bibinfo{author}{Hammond, R.~A.}
\newblock \bibinfo{journal}{\bibinfo{title}{Coupled contagion dynamics of fear
  and disease: mathematical and computational explorations}}.
\newblock {\emph{\JournalTitle{PloS one}}} \textbf{\bibinfo{volume}{3}},
  \bibinfo{pages}{e3955} (\bibinfo{year}{2008}).

\bibitem{epstein2021triple}
\bibinfo{author}{Epstein, J.~M.}, \bibinfo{author}{Hatna, E.} \&
  \bibinfo{author}{Crodelle, J.}
\newblock \bibinfo{journal}{\bibinfo{title}{Triple contagion: a two-fears
  epidemic model}}.
\newblock {\emph{\JournalTitle{Journal of the Royal Society Interface}}}
  \textbf{\bibinfo{volume}{18}}, \bibinfo{pages}{20210186}
  (\bibinfo{year}{2021}).

\bibitem{poletti2011effect}
\bibinfo{author}{Poletti, P.}, \bibinfo{author}{Ajelli, M.} \&
  \bibinfo{author}{Merler, S.}
\newblock \bibinfo{journal}{\bibinfo{title}{The effect of risk perception on
  the 2009 h1n1 pandemic influenza dynamics}}.
\newblock {\emph{\JournalTitle{PloS one}}} \textbf{\bibinfo{volume}{6}},
  \bibinfo{pages}{e16460} (\bibinfo{year}{2011}).

\bibitem{jovanovic2021modelling}
\bibinfo{author}{Jovanovi{\'c}, R.}, \bibinfo{author}{Davidovi{\'c}, M.},
  \bibinfo{author}{Lazovi{\'c}, I.}, \bibinfo{author}{Jovanovi{\'c}, M.} \&
  \bibinfo{author}{Jova{\v{s}}evi{\'c}-Stojanovi{\'c}, M.}
\newblock \bibinfo{journal}{\bibinfo{title}{Modelling voluntary general
  population vaccination strategies during covid-19 outbreak: influence of
  disease prevalence}}.
\newblock {\emph{\JournalTitle{International Journal of Environmental Research
  and Public Health}}} \textbf{\bibinfo{volume}{18}}, \bibinfo{pages}{6217}
  (\bibinfo{year}{2021}).

\bibitem{rajabi2021investigating}
\bibinfo{author}{Rajabi, A.}, \bibinfo{author}{Mantzaris, A.~V.},
  \bibinfo{author}{Mutlu, E.~C.} \& \bibinfo{author}{Garibay, O.~O.}
\newblock \bibinfo{journal}{\bibinfo{title}{Investigating dynamics of covid-19
  spread and containment with agent-based modeling}}.
\newblock {\emph{\JournalTitle{Applied Sciences}}}
  \textbf{\bibinfo{volume}{11}}, \bibinfo{pages}{5367} (\bibinfo{year}{2021}).

\bibitem{palomo2022agent}
\bibinfo{author}{Palomo-Briones, G.~A.}, \bibinfo{author}{Siller, M.} \&
  \bibinfo{author}{Grignard, A.}
\newblock \bibinfo{journal}{\bibinfo{title}{An agent-based model of the dual
  causality between individual and collective behaviors in an epidemic}}.
\newblock {\emph{\JournalTitle{Computers in biology and medicine}}}
  \textbf{\bibinfo{volume}{141}}, \bibinfo{pages}{104995}
  (\bibinfo{year}{2022}).

\bibitem{prieto2021vaccination}
\bibinfo{author}{Prieto~Curiel, R.} \&
  \bibinfo{author}{Gonz{\'a}lez~Ram{\'\i}rez, H.}
\newblock \bibinfo{journal}{\bibinfo{title}{Vaccination strategies against
  covid-19 and the diffusion of anti-vaccination views}}.
\newblock {\emph{\JournalTitle{Scientific Reports}}}
  \textbf{\bibinfo{volume}{11}}, \bibinfo{pages}{6626} (\bibinfo{year}{2021}).

\bibitem{tuccillo2023urbanpop}
\bibinfo{author}{Tuccillo, J.} \emph{et~al.}
\newblock \bibinfo{journal}{\bibinfo{title}{Urbanpop: A spatial microsimulation
  framework for exploring demographic influences on human dynamics}}.
\newblock {\emph{\JournalTitle{Applied Geography}}}
  \textbf{\bibinfo{volume}{151}}, \bibinfo{pages}{102844}
  (\bibinfo{year}{2023}).

\bibitem{pan2020quantifying}
\bibinfo{author}{Pan, Y.} \emph{et~al.}
\newblock \bibinfo{journal}{\bibinfo{title}{Quantifying human mobility
  behaviour changes during the covid-19 outbreak in the united states}}.
\newblock {\emph{\JournalTitle{Scientific Reports}}}
  \textbf{\bibinfo{volume}{10}}, \bibinfo{pages}{20742} (\bibinfo{year}{2020}).

\bibitem{delphi2022ctis}
\bibinfo{author}{{Carnegie Mellon University, Delphia Group}}.
\newblock \bibinfo{title}{{COVID-19 Trends and Impact Survey (CTIS) Results}}.
\newblock
  \bibinfo{howpublished}{\url{https://delphi.cmu.edu/covidcast/survey-results/?date=20220625##symptoms},
  last accessed Nov 8th, 2024} (\bibinfo{year}{2022}).

\bibitem{cdcPlanningScenarios2020}
\bibinfo{author}{{U.S. Centers for Disease Control and Prevention (CDC)}}.
\newblock
  \bibinfo{howpublished}{\url{https://archive.cdc.gov/www_cdc_gov/coronavirus/2019-ncov/hcp/planning-scenarios.html}}
  (\bibinfo{year}{2021}).

\end{thebibliography}

\supplementarysection{Supplementary Material}

\subsection{Coupled contagions ODE models}
\label{sup:odes}

\captionsetup[table]{name=Supplementary Table}
\captionsetup[figure]{name=Supplementary Figure}

We provide the full formulation of the system of ODEs which define the $SEPI_sI_aR_sR_a \times NF$ model below. This system is summarized in the flow diagram in Supplementary Fig.~\ref{fig:ode-flow-diagram}. Let $<x, y>$ be the size of the compartment with disease state $x$ and fear state $y$. The $SI_sI_aR_sR_a \times NF$ model can be obtained by removing all terms involving the \textbf{E}xposed and \textbf{P}resymptomatic infectious disease states (i.e. $<E, N>, <E, F>, <P, N>,$ and $<P, F>$). The $SIR \times NF$ model may similarly be obtained by additionally removing all terms involving \textbf{a}symptomatic disease states (i.e. $<I_a, N>, <I_a, F>, <R_a, N>,$ and $<R_a, F>$). The system of ODEs is as follows:

\begin{figure*}
    \centering
    \includegraphics[width=\linewidth]{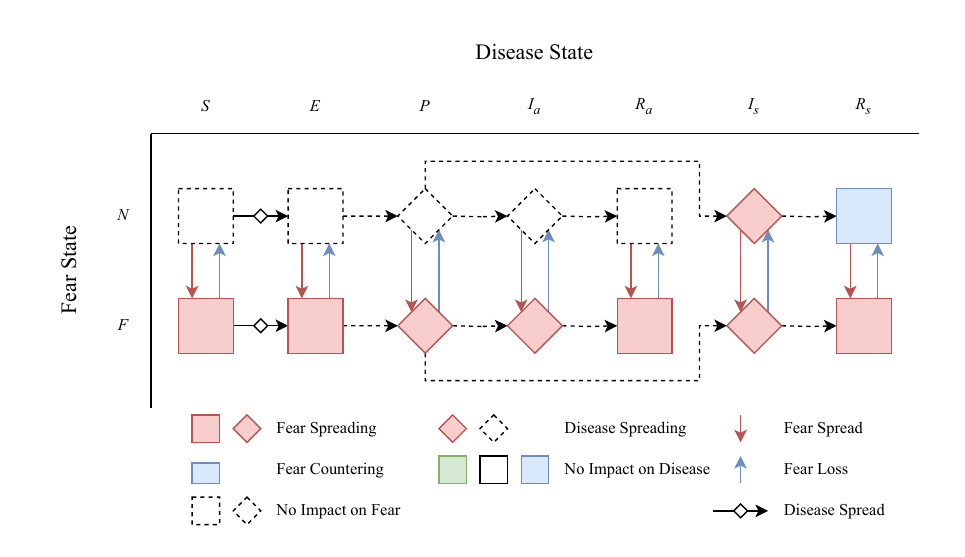}
    \caption{\bf Flow diagram for $SEPI_sI_aR_sR_a \times NF$ coupled contagion ODE model}
    \label{fig:ode-flow-diagram}
\end{figure*}

\begin{minipage}{.95\columnwidth}
\begin{align*}
    \text{fear}_{\uparrow}(<\bullet, N>) =&\; \eqnmarkbox[color_2]{beta_f}{\beta_f} \left( <I_s, N> + \sum_x <x, F> \right) <\bullet, N>\\
    \text{fear}_{\downarrow}(<\bullet, F>)  =&\; \left( \eqnmarkbox[color_5]{gamma_f}{\gamma_f} + \eqnmarkbox[color_3]{alpha_f}{\alpha_f} \cdot <R_s, N> \right) <\bullet, F>
\end{align*}
\annotate[yshift=.75em]{above}{beta_f}{\text{fear} transmission rate}
\annotate[yshift=-.15]{below,left}{gamma_f}{baseline \text{fear} loss rate}
\annotate[yshift=-.25]{below,right}{alpha_f}{\text{fear} loss rate from contact with symptomatic recovered}
\end{minipage}
\vspace{0.05in}

\begin{minipage}{.95\columnwidth}
\begin{align*}
    \text{disease}_{\uparrow}(<S, \bullet>) =&\; \eqnmarkbox[color_2]{beta}{\beta} ( <I_s, N> + \eqnmarkbox[color_4]{p1}{\iota_f} \cdot <I_s, F> \\
                               &+ \eqnmarkbox[color_6]{beta_a}{\iota_a} ( <P, N> + <I_a, N> \\
                               &+ \eqnmarkbox[color_4]{p2}{\iota_f} ( <P, F> + <I_a, F> ) ) ) <S, \bullet>
\end{align*}
\annotate[yshift=.5em]{above,left}{beta}{\text{disease} transmission rate}
\annotate[yshift=.65em,xshift=-1.5em]{below,left}{beta_a}{relative infectivity of asymptomatic individuals}
\annotate[yshift=.5em]{above,right}{p1}{relative infectivity of fearful individuals}
\end{minipage}
\vspace{-0.2in}

\begin{minipage}{.95\columnwidth}
\begin{align*}
    \text{disease}_{\downarrow}(<x, \bullet>) =&\; \eqnmarkbox[color_5]{gamma}{\gamma} \cdot <x, \bullet> : x \in \{I_s, I_a\}
\end{align*}
\annotate[yshift=.5em]{above}{gamma}{\text{disease} recovery rate}
\end{minipage}
\vspace{-0.35in}

\begin{minipage}{.95\columnwidth}
\begin{align*}
    \frac{d <S, N>}{dt} =&\; -\text{disease}_\uparrow(<S, N>) \\
    &- \text{fear}_\uparrow(<S, N>) + \text{fear}_\downarrow(<S, F>)
\end{align*}
\end{minipage}
\vspace{0.05in}

\begin{minipage}{.95\columnwidth}
\begin{align*}
    \frac{d <S, F>}{dt} =&\; -\eqnmarkbox[color_1]{sigma_f}{\sigma_f} \cdot \text{disease}_\uparrow(<S, F>) \\
    &+ \text{fear}_\uparrow(<S, N>) - \text{fear}_\downarrow(<S, F>)
\end{align*}
\annotate[yshift=.5em]{above}{sigma_f}{\text{disease} susceptibility reduction from \text{fear}}
\end{minipage}
\vspace{-0.2in}

\begin{minipage}{.95\columnwidth}
\begin{align*}
    \frac{d <E, N>}{dt} =&\; \text{disease}_\uparrow(<S, N>) - \eqnmarkbox[color_3]{delta1}{\delta} \cdot <E, N> \\
    &+ \text{fear}_{\uparrow}(<E, N>) + \text{fear}_{\downarrow}(<E, F>)\\
    \frac{d <E, F>}{dt} =&\; \eqnmarkbox[color_1]{sigma_f}{\sigma_f} \cdot \text{disease}_\uparrow(<S, F>) + \eqnmarkbox[color_3]{delta2}{\delta} \cdot <E, N>\\
    &- \text{fear}_{\uparrow}(<E, N>) + \text{fear}_{\downarrow}(<E, F>)\\
    \frac{d <P, N>}{dt} =&\; \eqnmarkbox[color_3]{delta2}{\delta} \cdot <E, N> - <P, N>\\
    \frac{d <P, F>}{dt} =&\; \eqnmarkbox[color_3]{delta2}{\delta} \cdot <E, F> - <P, F>
\end{align*}
\annotate[yshift=.5em]{above}{delta1}{inverse of incubation period}
\end{minipage}
\vspace{-0.2in}

\begin{minipage}{.95\columnwidth}
\begin{align*}
    \frac{d <I_s, N>}{dt} =&\; \eqnmarkbox[color_4]{p_s1}{p_s} \cdot <P, N> - \text{disease}_{\downarrow}(<I_s, N>)\\
    &- \text{fear}_{\uparrow}(<I_s, N>) + \text{fear}_{\uparrow}(<I_s, F>) \\
    \frac{d <I_s, F>}{dt} =&\; \eqnmarkbox[color_4]{p_s2}{p_s} \cdot <P, N> - \text{disease}_{\downarrow}(<I_s, N>)\\
    &+ \text{fear}_{\uparrow}(<I_s, N>) - \text{fear}_{\uparrow}(<I_s, F>)
\end{align*}
\annotate[yshift=.5em]{above}{p_s1}{probability of symptomatic infection}
\end{minipage}
\vspace{-0.2in}

\begin{minipage}{.95\columnwidth}
\begin{align*}
    \frac{d <I_a, N>}{dt} =&\; \eqnmarkbox[color_5]{p_a1}{p_a} \cdot <P, N> - \text{disease}_{\downarrow}(<I_a, N>)\\
    &- \text{fear}_{\uparrow}(<I_a, N>) + \text{fear}_{\uparrow}(<I_a, F>) \\
    \frac{d <I_a, F>}{dt} =&\; \eqnmarkbox[color_5]{p_a2}{p_a} \cdot <P, F> - \text{disease}_{\downarrow}(<I_a, N>)\\
    &+ \text{fear}_{\uparrow}(<I_a, N>) - \text{fear}_{\uparrow}(<I_a, F>)
\end{align*}
\annotate[yshift=.5em]{above}{p_a1}{probability of asymptomatic infection}
\end{minipage}

\begin{minipage}{.95\columnwidth}
\begin{align*}
    \frac{d <R_s, N>}{dt} =&\; \text{disease}_{\downarrow}(<I_s, N>) \\
    &- \eqnmarkbox[color_2]{rho1}{\rho_f} \cdot \text{fear}_{\uparrow}(<R_s, N>) + \text{fear}_{\uparrow}(<R_s, F>) \\
    \frac{d <R_s, F>}{dt} =&\; \text{disease}_{\downarrow}(<I_s, F>)\\
    &+ \eqnmarkbox[color_2]{rho2}{\rho_f} \cdot \text{fear}_{\uparrow}(<R_s, F>) - \text{fear}_{\uparrow}(<R_s, F>)
\end{align*}
\annotate[yshift=-.25em]{below}{rho2}{\text{fear} susceptibility reduction from symptomatic recovery}
\end{minipage}
\vspace{-0.15in}

\begin{minipage}{.95\columnwidth}
\begin{align*}
    \frac{d <R_a, N>}{dt} =&\; \text{disease}_{\downarrow}(<I_a, N>)\\
    &- \text{fear}_{\uparrow}(<R_a, N>) + \text{fear}_{\uparrow}(<R_a, F>) \\
    \frac{d <R_a, F>}{dt} =&\; \text{disease}_{\downarrow}(<I_a, F>)\\
    &+ \text{fear}_{\uparrow}(<R_a, F>) - \text{fear}_{\uparrow}(<R_a, F>)
\end{align*}
\end{minipage}

\subsection{Model parameters}
\label{sup:params}

Parameter values used in ODE model comparison experiments (Section~4.1) are shown in Supplementary Table~\ref{tab:ode-params} while those used in the EpiCast scenario comparison experiments (Section~4.2) are shown in Supplementary Table~\ref{tab:scenario-params}. \textcolor{revisions}{Where possible, sources for the values used are provided. For rows in which multiple values are used, the values without a citation are generally chosen to contrast with the cited values}.

\begin{table}[!ht]
\centering
\begin{tabular}{@{}rrrrrrrr@{}}
\toprule
\multicolumn{1}{c}{Variable} & \multicolumn{1}{c}{Description} & \multicolumn{6}{c}{Experiment} \\
\multicolumn{1}{c}{} & \multicolumn{1}{c}{} & \multicolumn{1}{c}{(a)} & \multicolumn{1}{c}{(b)} & \multicolumn{1}{c}{(c)} & \multicolumn{1}{c}{(d)} & \multicolumn{1}{c}{(e)} & \multicolumn{1}{c}{(f)} \\ \midrule
$\beta$ & disease transmission rate & 0.2 & 0.2 & 0.2 & 0.2 & 0.2 & 0.2 \\
$\gamma$ & disease recovery rate & 0.5 & 0.5 & 0.5 & 0.5 & 0.5 & 0.5 \\
$p_s$ & \begin{tabular}[c]{@{}r@{}}probability of\\ symptomatic infection\end{tabular} &  &  & 0.6~\textcolor{revisions}{\cite{cdcPlanningScenarios2020}} & 0.6 & 0.6 & 0.6 \\
$p_a = 1 - p_s$ & \begin{tabular}[c]{@{}r@{}}probability of \\ asymptomatic infection\end{tabular} &  &  & 0.4~\textcolor{revisions}{\cite{cdcPlanningScenarios2020}} & 0.4 & 0.4 & 0.4 \\
$\iota_a$ & \begin{tabular}[c]{@{}r@{}}relative infectivity of\\ asymptomatic infection\end{tabular} &  &  & 0.75~\textcolor{revisions}{\cite{cdcPlanningScenarios2020}} & 0.75 & 0.75 & 0.75 \\
$\delta$ & \begin{tabular}[c]{@{}r@{}}inverse of\\ incubation period\end{tabular} &  &  &  &  & 0.5 & 0.5 \\ \midrule
$\beta_f$ & \begin{tabular}[c]{@{}r@{}}fear transmission\\ rate\end{tabular} & \textcolor{revisions}{$1.1\beta$~\cite{epstein2021triple}} & \textcolor{revisions}{$1.1\beta$} & \textcolor{revisions}{$1.1\beta$} & \textcolor{revisions}{$1.1\beta$} & \textcolor{revisions}{$1.1\beta$} & \textcolor{revisions}{$1.1\beta$} \\
$\gamma_f$ & baseline fear loss rate & 0.05~\cite{epstein2021triple} & 0.05 & 0.05 & 0.05 & 0.05 & 0.05 \\
$\alpha_f$ & fear loss contact rate & \textcolor{revisions}{$2.2\beta$~\cite{epstein2021triple}} & \textcolor{revisions}{$2.2\beta$} & \textcolor{revisions}{$2.2\beta$} & \textcolor{revisions}{$2.2\beta$} & \textcolor{revisions}{$2.2\beta$} & \textcolor{revisions}{$2.2\beta$} \\
$\rho_f$ & \begin{tabular}[c]{@{}r@{}}relative fear\\ susceptibility after\\ symptomatic recovery\end{tabular} & 1 & 0~\cite{epstein2021triple} & 0 & 0 & 0 & 0 \\ \midrule
$\iota_f$ & \begin{tabular}[c]{@{}r@{}}relative infectivity\\ when fearful\end{tabular} & 1~\textcolor{revisions}{\cite{cdcPlanningScenarios2020}} & 1 & 1 & 1 & 1 & 1 \\
$\sigma_f$ & \begin{tabular}[c]{@{}r@{}}relative susceptibility\\ when fearful\end{tabular} & 0.25~\textcolor{revisions}{\cite{epstein2021triple}} & 0.25 & 0.25 & 0.35 & 0.25 & 0.35 \\ \bottomrule
\end{tabular}
\caption{{\bf Parameter values used in ODE model experiments:} Results of experiments (a-e) are shown in the corresponding subfigures in Fig.~1, and the EpiCast scenarios to setups with (a) only hospitalization-based withdrawals, (b) the addition of symptomatic fearful withdrawals, and the addition of either pure-fear withdrawals from fear -- without (c) or with (d) broadcaster-based fear spread -- or reduced susceptibility when fearful -- without (e) or with (f) broadcaster-based fear spread.}
\label{tab:ode-params}
\end{table}

\begin{table}[!ht]
\centering
\begin{tabular}{rrrrrrrr}
\toprule
\multicolumn{1}{c}{Variable} & \multicolumn{1}{c}{Description} & \multicolumn{6}{c}{Scenario} \\
\multicolumn{1}{c}{} & \multicolumn{1}{c}{} & \multicolumn{1}{c}{(a)} & \multicolumn{1}{c}{(b)} & \multicolumn{1}{c}{(c)} & \multicolumn{1}{c}{(d)} & \multicolumn{1}{c}{(e)} & \multicolumn{1}{c}{(f)} \\ \midrule
$\beta$ & disease transmission rate & 0.2 & 0.2 & 0.2 & 0.2 & 0.2 & 0.2 \\
$\gamma$ & disease recovery rate & 0.5 & 0.5 & 0.5 & 0.5 & 0.5 & 0.5 \\
$p_s$ & \begin{tabular}[c]{@{}r@{}}probability of\\ symptomatic infection\end{tabular} & 0.6 & 0.6 & 0.6 & 0.6 & 0.6 & 0.6~\textcolor{revisions}{\cite{cdcPlanningScenarios2020}} \\
$p_a = 1 - p_s$ & \begin{tabular}[c]{@{}r@{}}probability of \\ asymptomatic infection\end{tabular} & 0.4 & 0.4 & 0.4 & 0.4 & 0.4 & 0.4~\textcolor{revisions}{\cite{cdcPlanningScenarios2020}} \\
$\iota_a$ & \begin{tabular}[c]{@{}r@{}}relative infectivity of\\ asymptomatic infection\end{tabular} & 0.75 & 0.75 & 0.75 & 0.75 & 0.75 & 0.75~\textcolor{revisions}{\cite{cdcPlanningScenarios2020}} \\
$\delta$ & \begin{tabular}[c]{@{}r@{}}inverse of\\ incubation period\end{tabular} & 0.5 & 0.5 & 0.5 & 0.5 & 0.5 & 0.5  \\ \midrule
$\beta_f$ & \begin{tabular}[c]{@{}r@{}}fear transmission\\ rate\end{tabular} & \textcolor{revisions}{$1.1\beta$} & \textcolor{revisions}{$1.1\beta$} & \textcolor{revisions}{$1.1\beta$} & \textcolor{revisions}{$1.1\beta$} & \textcolor{revisions}{$1.1\beta$} & \textcolor{revisions}{$1.1\beta$~\cite{epstein2021triple}} \\
$\gamma_f$ & baseline fear loss rate & 0.05 & 0.05 & 0.05 & 0.05 & 0.05 & 0.05~\textcolor{revisions}{\cite{epstein2021triple}} \\
$\alpha_f$ & fear loss contact rate & \textcolor{revisions}{$2.2\beta$} & \textcolor{revisions}{$2.2\beta$} & \textcolor{revisions}{$2.2\beta$} & \textcolor{revisions}{$2.2\beta$} & \textcolor{revisions}{$2.2\beta$} & \textcolor{revisions}{$2.2\beta$~\cite{epstein2021triple}} \\
$\rho_f$ & \begin{tabular}[c]{@{}r@{}}relative fear\\ susceptibility after\\ symptomatic recovery\end{tabular} & 0 & 0 & 0 & 0 & 0 & 0~\textcolor{revisions}{\cite{epstein2021triple}} \\ \midrule
$\iota_f$ & \begin{tabular}[c]{@{}r@{}}relative infectivity\\ when fearful\end{tabular} & 1~\textcolor{revisions}{\cite{cdcPlanningScenarios2020}} & 1 & 1 & 1 & 1 & 1 \\
$\sigma_f$ & \begin{tabular}[c]{@{}r@{}}relative susceptibility\\ when fearful\end{tabular} & 1 & 1 & 1 & 1 & 0.35 & 0.35 \\
$p_\text{sick}$ & \begin{tabular}[c]{@{}r@{}}probability of withdrawal\\ when fearful with symptoms\end{tabular} & 0 & 1 & 1 & 1 & 1 & 1 \\
$p_\text{fear}$ & \begin{tabular}[c]{@{}r@{}}probability of pure-fear\\ withdrawal when fearful\end{tabular} & 0 & 0 & 0.65 & 0.65 & 0 & 0 \\ \midrule
$p_\text{bc}$ & \begin{tabular}[c]{@{}r@{}}probability of watching\\ a given broadcaster\end{tabular} & 0 & 0 & 0 & 0.25 & 0 & 0.25 \\
$p_\text{bc\_start}$ & \begin{tabular}[c]{@{}r@{}}threshold of fearful\\ workers for broadcasters\\ to take position\end{tabular} &  &  &  & 0.5 &  & 0.5 \\
$p_\text{bc\_neutral}$ & \begin{tabular}[c]{@{}r@{}}relative rate of new cases\\ for broadcasters to\\ resume neutral position\end{tabular} &  &  &  & \textcolor{revisions}{$\frac{p_\text{bc\_start}}{2}$} &  & \textcolor{revisions}{$\frac{p_\text{bc\_start}}{2}$} \\
$p_\text{bc\_counter}$ & \begin{tabular}[c]{@{}r@{}}relative rate of new cases\\ for broadcasters\\ to counter fear spread\end{tabular} &  &  &  & \textcolor{revisions}{$\frac{p_\text{bc\_start}}{4}$} &  & \textcolor{revisions}{$\frac{p_\text{bc\_start}}{4}$} \\ \bottomrule
\end{tabular}

\caption{{\bf Parameter values used in EpiCast scenario experiments:} Parameters represent scenarios with (a) only hospitalization-based withdrawals, (b) the addition of symptomatic fearful withdrawals, and the addition of either pure-fear withdrawals -- without (c) or with (d) broadcaster-based fear spread -- or reduced susceptibility when fearful -- without (e) or with (f) broadcaster-based fear spread.}
\label{tab:scenario-params}
\end{table}

{\color{revisions}
\subsection{Additional Sensitivity Analysis}

We conduct another set of experiments which vary the distribution of initial cases used to seed infections in EpiCast, along with the values of $p_\text{bc\_start}$, using several different random seeds for each cell. For these runs, we use the population of the U.S. state of Colorado ($\sim 5.6$ million agents) with three different starting conditions:
\begin{enumerate}
    \item March 25: Infections and immune agents are seeded with case data from March 25 2020, similar to the other EpiCast runs presented above.
    \item Denver only: Infections and immune agents are seeded only in Denver County.
    \item All counties: Infections and immune agents are seeded in equal number in every county of Colorado. Counts are not scaled by the population of the county.
\end{enumerate}
Note that in all cases we seed approximately the same number of infections and immune agents, though in the final case we round down the per-county averages. In all cases we use the parameter values from Scenario (f) except when otherwise specified.

We run two sets of simulations. In the first, shown in Fig.~\ref{fig:sa-ic-1}, we use $\sigma_f = 0$ and $p_\text{fear} = 0.25$ to test a set of parameters that produced a single wave for the state of Colorado experiments. In the second, shown in Fig.~\ref{fig:sa-ic-2} we use $\sigma_f = 0.5$ and $p_\text{fear} = 0$ to test a set of parameters that produces two waves.

For the first set of runs, we observe relatively little variation based on random seed within a given cell. We also note that larger values of $p_\text{bc\_start}$ as much as doubles the peak rate of new infections. With respect to initial conditions, seeding based on the observed cases from March 25 produces the most infections, given the same value of $p_\text{bc\_start}$. All cases produce one wave.

For the second set of runs all cases produce two waves. Here we observe greater variability for $p_\text{bc\_start} = 0.75$, along with much greater decay in fear levels from their initial peak and a corresponding heightened second peak of the epidemic. Differences in cases due to initial conditions are relatively minor in this regime.

\begin{figure*}
    \centering
    \includegraphics[width=\linewidth]{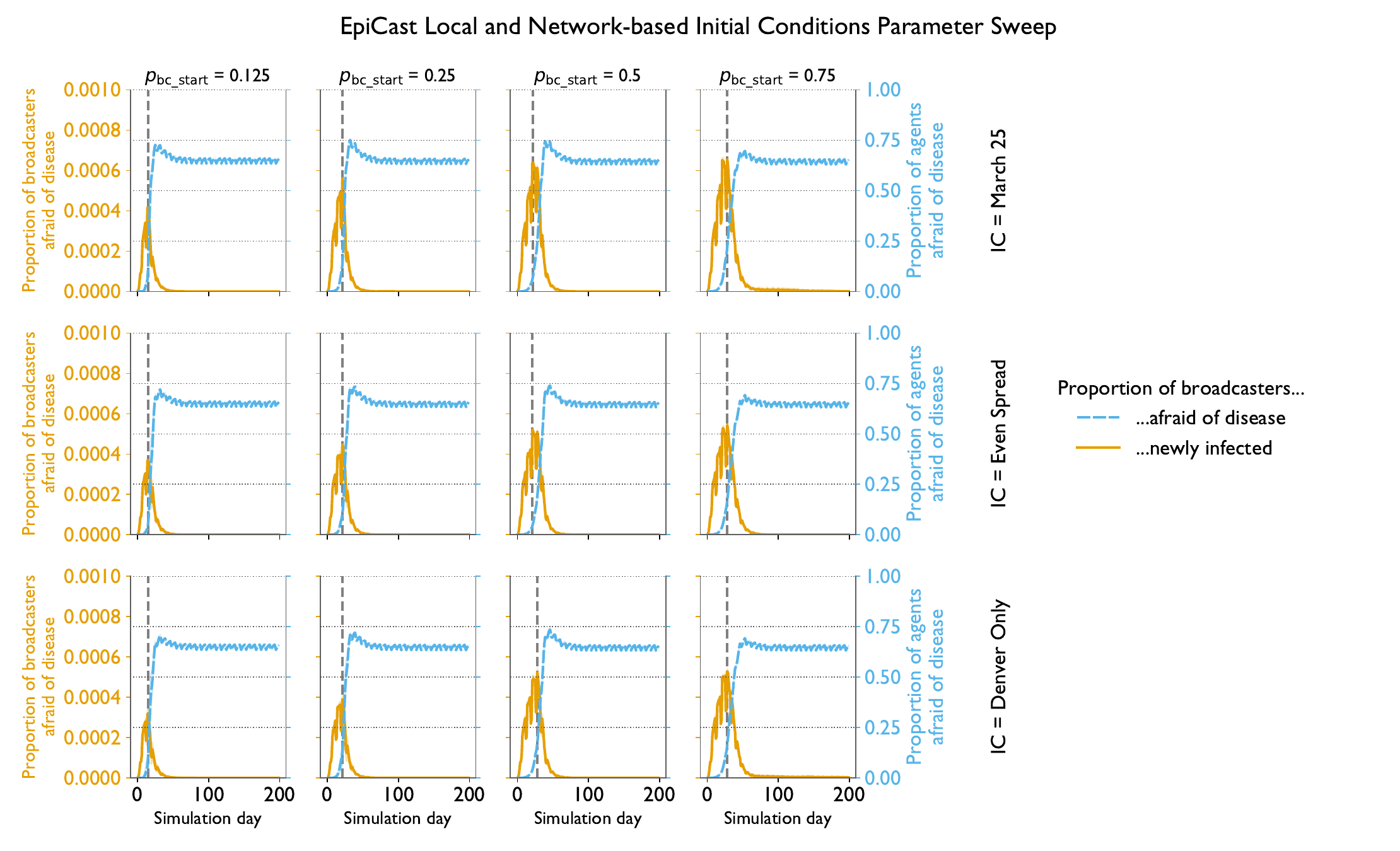}
    \caption{\textcolor{revisions}{\textbf{Epicast one-wave initial condition and broadcaster threshold sensitivity analysis:} New cases by day (with peaks indicated by dashed vertical line), for various initial conditions (IC) and thresholds for broadcasters to start spreading fear ($p_\text{bc\_start}$) with both local and broadcaster-based fear spread. All lines represent the average of three replicates, with a 95\% confidence interval shown in shading.}}
    \label{fig:sa-ic-1}
\end{figure*}

\begin{figure*}
    \centering
    \includegraphics[width=\linewidth]{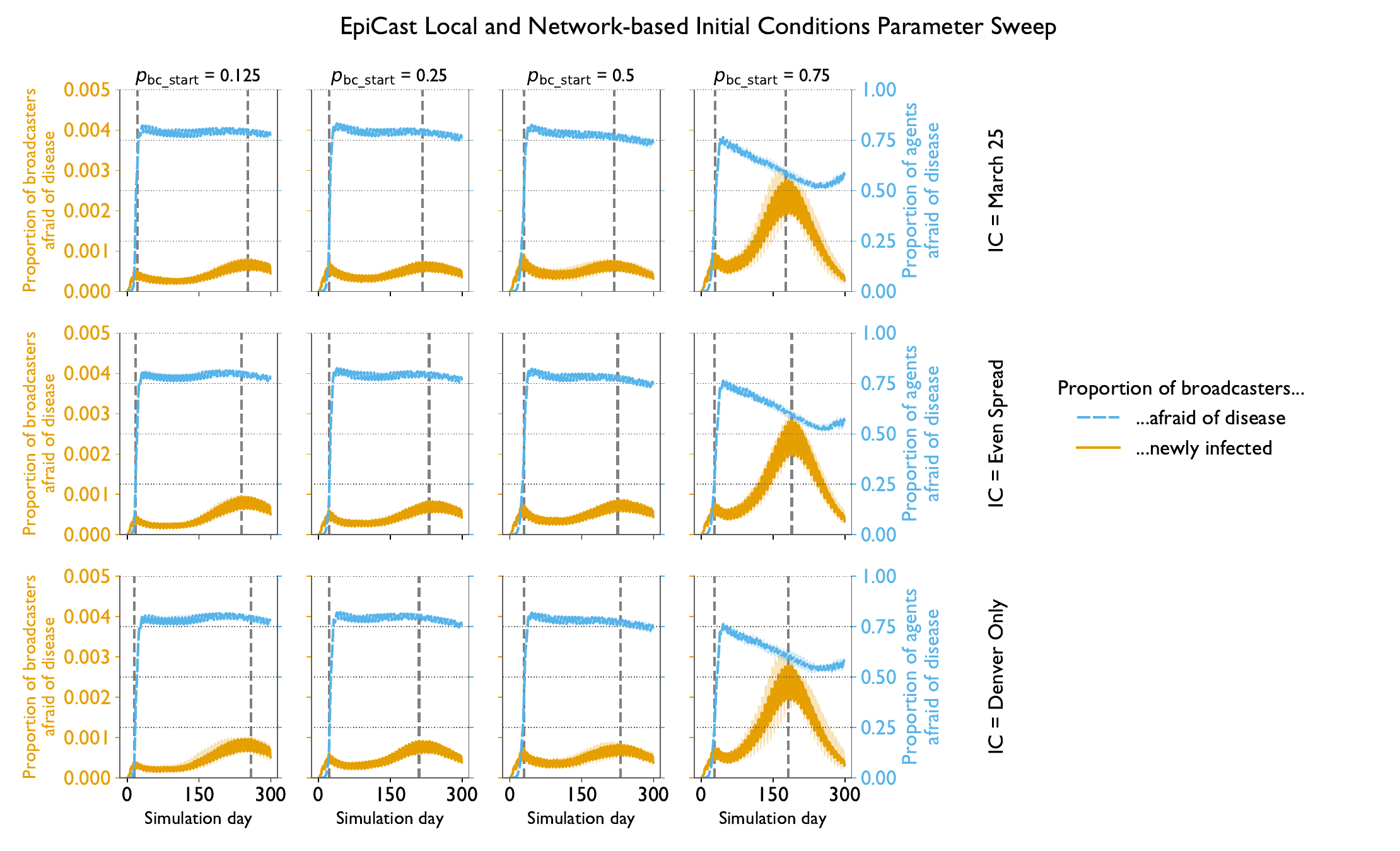}
    \caption{\textcolor{revisions}{\textbf{Epicast  two-wave initial condition and broadcaster threshold sensitivity analysis:} New cases by day (with peaks indicated by dashed vertical line), for various initial conditions (IC) and thresholds for broadcasters to start spreading fear ($p_\text{bc\_start}$) with both local and broadcaster-based fear spread. All lines represent the average of three replicates, with a 95\% confidence interval shown in shading.}}
    \label{fig:sa-ic-2}
\end{figure*}
}

\end{document}